\documentclass[fleqn,usenatbib]{mnras}

\usepackage{newtxtext,newtxmath}
\usepackage{graphicx}
\usepackage{bm}
\usepackage{enumitem}
\usepackage{makecell}
\usepackage{lineno}
\hypersetup{
	colorlinks = true,
	linkcolor = blue,
	anchorcolor = blue,
	citecolor = blue,
	urlcolor = blue
}
\usepackage[caption=false]{subfig}
\usepackage{soul, color, xcolor}
\usepackage{multirow}
\usepackage{booktabs}
\usepackage{threeparttable}

\usepackage[T1]{fontenc}

\DeclareRobustCommand{\VAN}[3]{#2}
\let\VANthebibliography\thebibliography
\def\thebibliography{\DeclareRobustCommand{\VAN}[3]{##3}\VANthebibliography}

\usepackage{graphicx}	
\usepackage{amsmath}	

\title[The X-ray re-brightening of GRB afterglow revisited]{The X-ray re-brightening of GRB afterglow revisited: a possible signature from activity of the central engine}
\author[Yang et al.]
{Zhe Yang$^{1}$, Hou-Jun L\"{u}$^{1}$\thanks{E-mail: lhj@gxu.edu.cn}, Xing Yang$^{1}$, Jun Shen$^{2}$, and
Shuang-Xi Yi$^{3}$  \\
 $^1$Guangxi Key Laboratory for Relativistic Astrophysics, School of Physical Science and Technology,
 Guangxi University, Nanning 530004, China\\
 $^2$Institute of Astrophysics, Central China Normal University, Wuhan 430079, China\\
 $^3$School of Physics and Physical Engineering, Qufu Normal University, Qufu 273165, China\\}

\date{Accepted XXX. Received YYY; in original form ZZZ}
\pubyear{2023}
\begin{document}
\maketitle
\begin{abstract}
Long-duration gamma-ray bursts (GRBs) are thought to be from core collapse of massive stars, and a rapidly spinning magnetar or black hole may be formed as the central engine. The extended emission in the prompt emission, flares and plateaus in X-ray afterglow, are proposed to be as the signature of central engine re-activity. However, the directly evidence from observations of identifying the central engines remain an open question. In this paper, we systemically search for long-duration GRBs that consist of bumps in X-ray afterglow detected by Swift/XRT, and find that the peak time of the X-ray bumps exhibit bimodal distribution (defined as “early” and “late” bumps) with division line at $t=7190$ s. Although we cannot rule out that such a bimodality arises from selection effects. We proposed that the long-duration GRBs with an early (or late) bumps may be originated from the fall-back accretion onto a new-born magnetar (or black hole). By adopting MCMC method to fit the early (or late) bumps of X-ray afterglow with the fall-back accretion of magnetar (or black hole), it is found that the initial surface magnetic filed and period of magnetars for most early bumps are clustered around $5.88\times10^{13}$ G and $1.04$ ms, respectively. Meanwhile, the derived accretion mass of black hole for late bumps is range of $[4\times10^{-4}, 1.8\times10^{-2}]~M_{\odot}$, and the typical fall-back radius is distributed range of $[1.04, 4.23]\times 10^{11}$ cm which is consistent with the typical radius of a Wolf-Rayet star. However, we also find that the fall-back accretion magnetar model is disfavored by the late bumps, but the fall-back accretion of black hole model can not be ruled out to interpret the early bumps of X-ray afterglow.
\end{abstract}


\section {Introduction}
In general, it is believed that gamma-ray bursts (GRBs) are originated from the collapsar of massive stars or the merger of double compact stars \citep{1989Natur.340..126E, 1993ApJ...405..273W, 1999ApJ...524..262M, 2018pgrb.book.....Z}. Within both the collapsar and compact-star merger models, a hyper-accretion black hole or a rapidly spinning, strongly magnetized neutron star (millisecond magnetar) as the central engine may be formed, and it launches a relativistic outflow \citep{1992Natur.357..472U, 1998PhRvL..81.4301D, 1998A&A...333L..87D, 1999ApJ...518..356P, 2001ApJ...557..949N, 2001ApJ...552L..35Z, 2006MNRAS.372L..19F, 2007ApJ...657..383C, 2010MNRAS.406.2650M, 2013ApJ...765..125L, 2014ApJ...785...74L, 2015ApJ...805...89L}. The observed prompt gamma-ray emission is explained by the fireball internal shocks \citep{1994ApJ...430L..93R}, dissipated photosphere models \citep{1994MNRAS.270..480T, 2005ApJ...628..847R, 2006ApJ...642..995P}, and the internal-collision-induced magnetic re-connection and turbulence (ICMART) model \citep{2011ApJ...726...90Z}. The broadband afterglow emission is produced from forward and reverse external shocks when the fireball is decelerated by a circumburst medium \citep{1997ApJ...476..232M, 1998ApJ...497L..17S, 2000ApJ...545..807K, 2002ARA&A..40..137M, 2004IJMPA..19.2385Z, 2016SSRv..202....3Z}. 

However, how to identify the central engine (black hole or magnetar) of GRBs remain an open question \citep{2011CRPhy..12..206Z}. From the observational point of view, some GRBs have been discovered to exhibit a plateau emission component or an extremely steep drop following the plateau (known as internal plateaus) in their X-ray afterglows \citep{2006ApJ...642..354Z, 2006ApJ...647.1213O, 2007ApJ...670..565L, 2007ApJ...665..599T, 2010MNRAS.402..705L, 2010MNRAS.409..531R, 2013MNRAS.430.1061R, 2014ApJ...785...74L, 2015ApJ...805...89L}, which is consistent with millisecond magnetar central engine \citep{1992Natur.357..472U, 1998PhRvL..81.4301D, 1998A&A...333L..87D, 2001ApJ...552L..35Z}. On the other hand, non-plateau emission in the afterglows, the released energy of GRB exceeded the energy budget of magnetar, or the later giant bump in the afterglows, are inconsistent with the magnetar central, but can be interpreted by black hole central engine, such as GRBs 121027A and 111209A \citep{2008Sci...321..376K, 2013ApJ...767L..36W, 2015ApJ...805...88Y, 2018MNRAS.480.4402L, 2021ApJ...906...60Z}.

From the theoretical point of view, \cite{2008ApJ...683..329Z, 2009ApJ...703..461Z} invoked a new-born neutron star surrounded by hyper-accretion and neutrino cooling disc to produce both GRB jet and observed plateau emission. Within this scenario, \cite{2012ApJ...759...58D} proposed a hyper-accretion fall-back disk around a newborn millisecond magnetar model to produce a significant brightening of an early afterglow (early X-ray bump) when magnetar is spin-up due to a sufficient angular momentum of the accreted matter transferred to the magnetar. If the central engine is black hole, a giant X-ray or optical bump can be produced via the fall-back of black hole when the duration and accretion rate of black hole fallback are long and large enough, respectively \citep{2013ApJ...767L..36W, 2015ApJ...805...88Y, 2018MNRAS.480.4402L, 2021ApJ...906...60Z}. If this is the case, different types of central engine of GRBs maybe produce different characteristic of X-ray or optical afterglows.

In this paper, we systematically search for long-duration GRBs with an early and later X-ray bumps emission from the Swift X-Ray Telescope (XRT) GRB sample, and try to interpret both early and later X-ray bumps by invoking fall-back accretion of magnetar and black hole, respectively. The criteria of sample selection and the data reduction are presented in section 2. In section 3, we described the basic models of both spin-up of magnetar fall-back and the black hole fall-back accretion. We apply the two models to fit the early and later X-ray afterglow of GRBs, respectively in section 4. Our conclusions and discussion are given in section 5. Throughout the paper, a concordance cosmology with parameters $H_0=71~\rm km~s^{-1}~Mpc^{-1}$, $\Omega_M=0.30$, and $\Omega_{\Lambda}=0.70$ is adopted.

\section{Data Reduction and Sample Selection Criteria}
The XRT data are downloaded from the Swift data archive \citep{2007A&A...469..379E}\footnote{http://www.swift.ac.uk/archive/obs.php?burst=1.}. Our entire sample includes more than 1718 GRBs observed by Swift/XRT between 2005 January and 2023 July. We only focus on the long-duration GRBs with the bump emission in X-ray afterglow, and analyze 1043 long-duration GRBs. Among these, 154 GRBs are too faint to be detected in the X-ray band, or do not have enough photons to extract a reasonable X-ray light curve. Then, we select the GRBs that X-ray emission exhibit the feature of rise and fall, and adopt a smooth broken power-law function to fit \citep{2007ApJ...670..565L,2022ApJ...931L..23L},
\begin{eqnarray}
F_1(t)=F_{01}\left [
\left (\frac{t}{t_p}\right)^{\omega\alpha_1}+\left(
\frac{t}{t_p}\right)^{\omega\alpha_2}\right]^{-1/\omega},
\label{SBPL}
\end{eqnarray}

\begin{eqnarray}
F_2(t)=\left (
F_1^{-\omega_1}+
F_3^{-\omega_1}\right)^{-1/\omega_1},
\label{SBPL}
\end{eqnarray}

\begin{eqnarray}
F_3(t)=F_{2}({t_{b,2}})
\left (\frac{t}{t_{b,2}}\right)^{-\alpha_3}
\label{SBPL}
\end{eqnarray}

where fixed $\omega=\omega_1=3$ represents the sharpness of the peak and $t_p$, $\alpha_1$, and $\alpha_2$
are the peak time, and the rising and decay slopes of X-ray variability, respectively.

Since the X-ray flares have been discovered in a good fraction of Swift GRBs \citep{2005SSRv..120..165B, 2007ApJ...671.1903C, 2010MNRAS.406.2149M}. Their lightcurves are typically narrow, and show rapid rise and fall with steep rising and decaying indices. This feature of X-ray flares are similar to that of prompt emission, and suggests that the X-ray flares are likely to share a similar mechanism with the prompt emission, such as internal dissipation of long-lasting central engine activity \citep{2005SSRv..120..165B, 2006ApJ...646..351L, 2006ApJ...642..354Z, 2006ApJ...642..389N, 2015ApJ...803...10T, 2016ApJS..224...20Y}. \cite{2016ApJS..224...20Y} performed a systematic study of X-ray flares observed by Swift to obtain the rising and decaying indices, and found that the rising and decaying indices of those flares are larger or much larger that 3. While, we find that a small fraction of GRBs x-ray afterglow whose rising and decaying indices are less than 3 (called X-ray bump). It means that the physical origin of those X-ray bumps may be different from that of X-ray flares, and it is possible related to the central central engine, e.g., spin-up of magnetar or fall-back of black hole.

\begin{figure*}
    \includegraphics[width=3.5in]{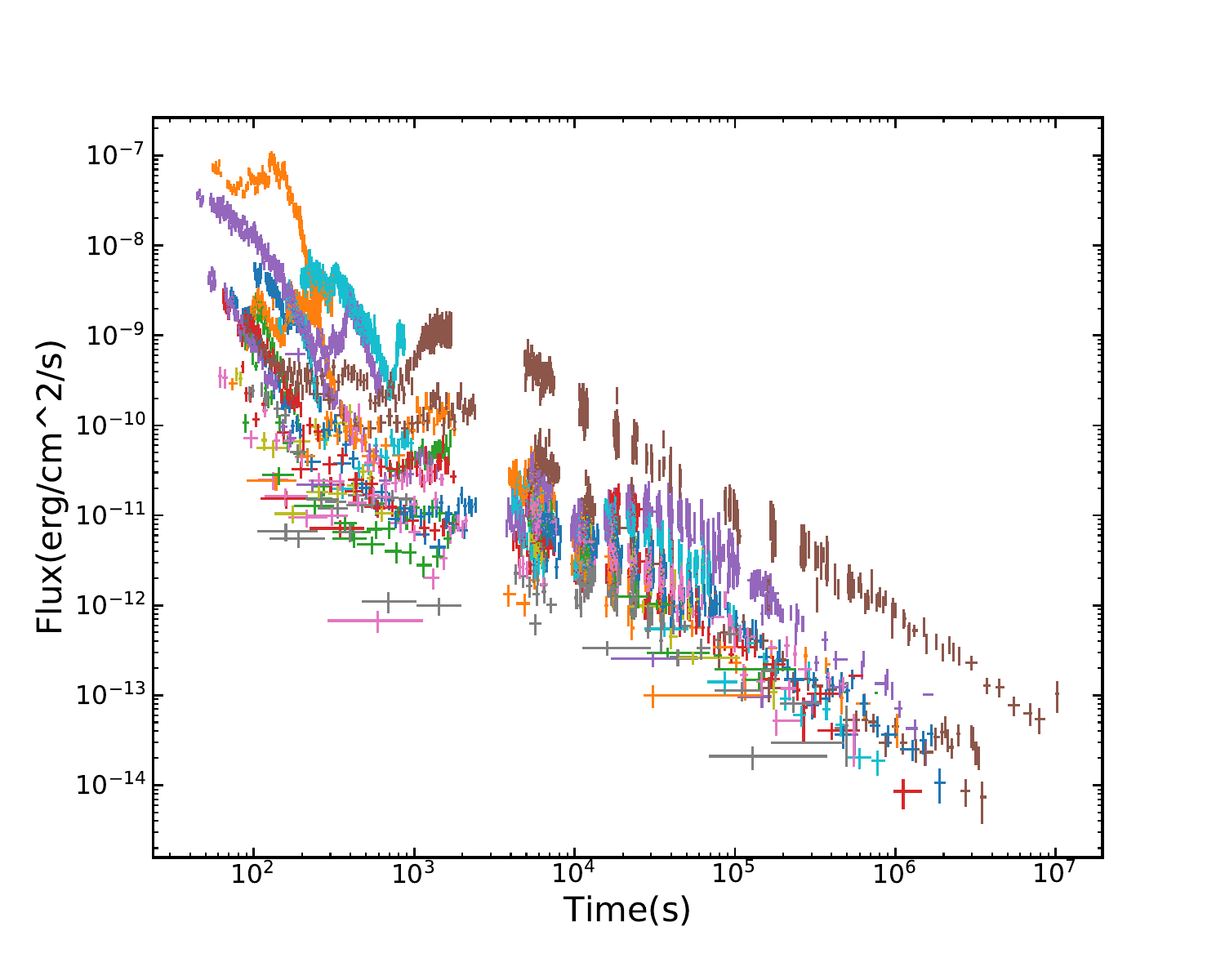}
    \includegraphics[width=3.5in]{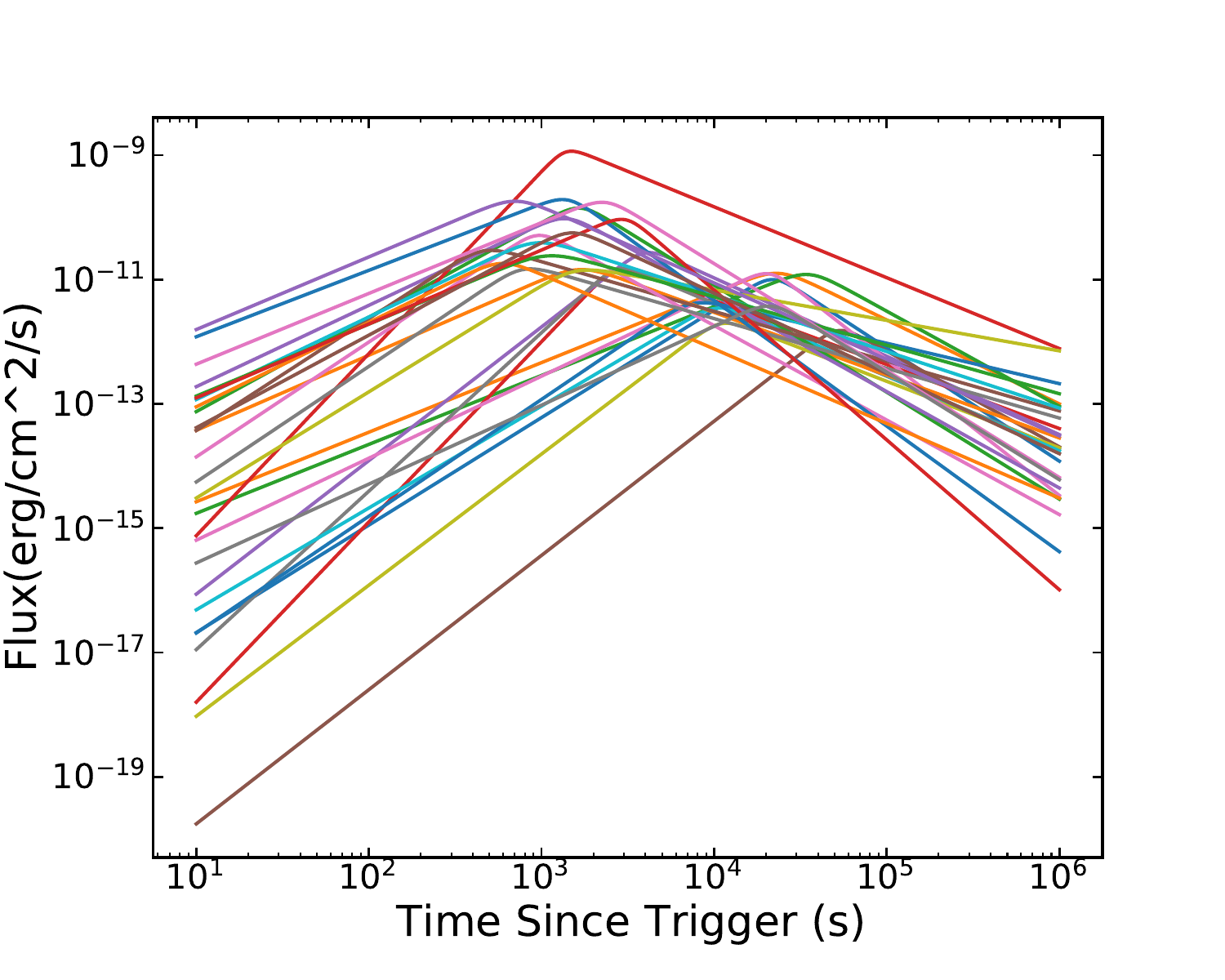}
    \includegraphics[width=3.5in]{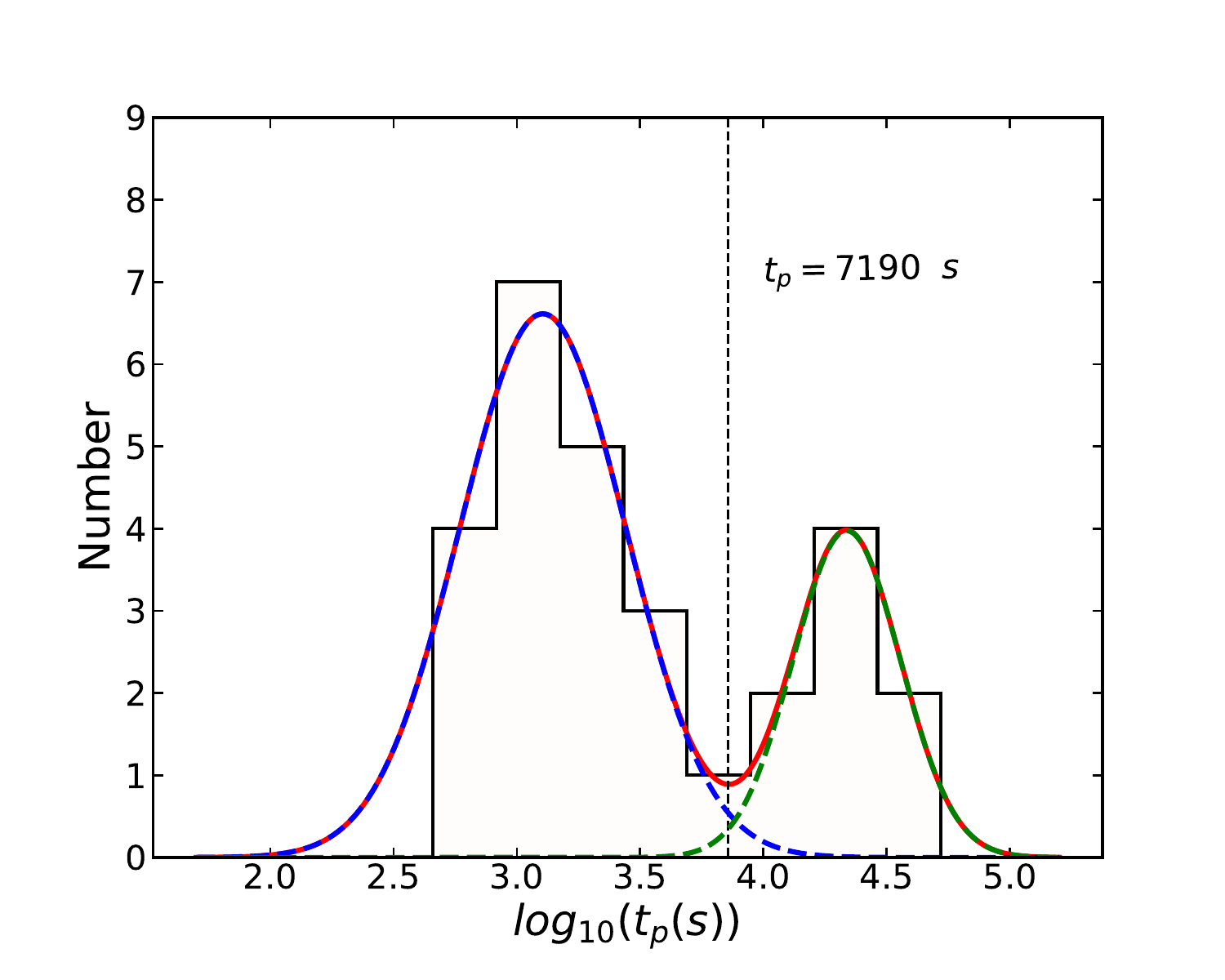}
    \includegraphics[width=3.5in]{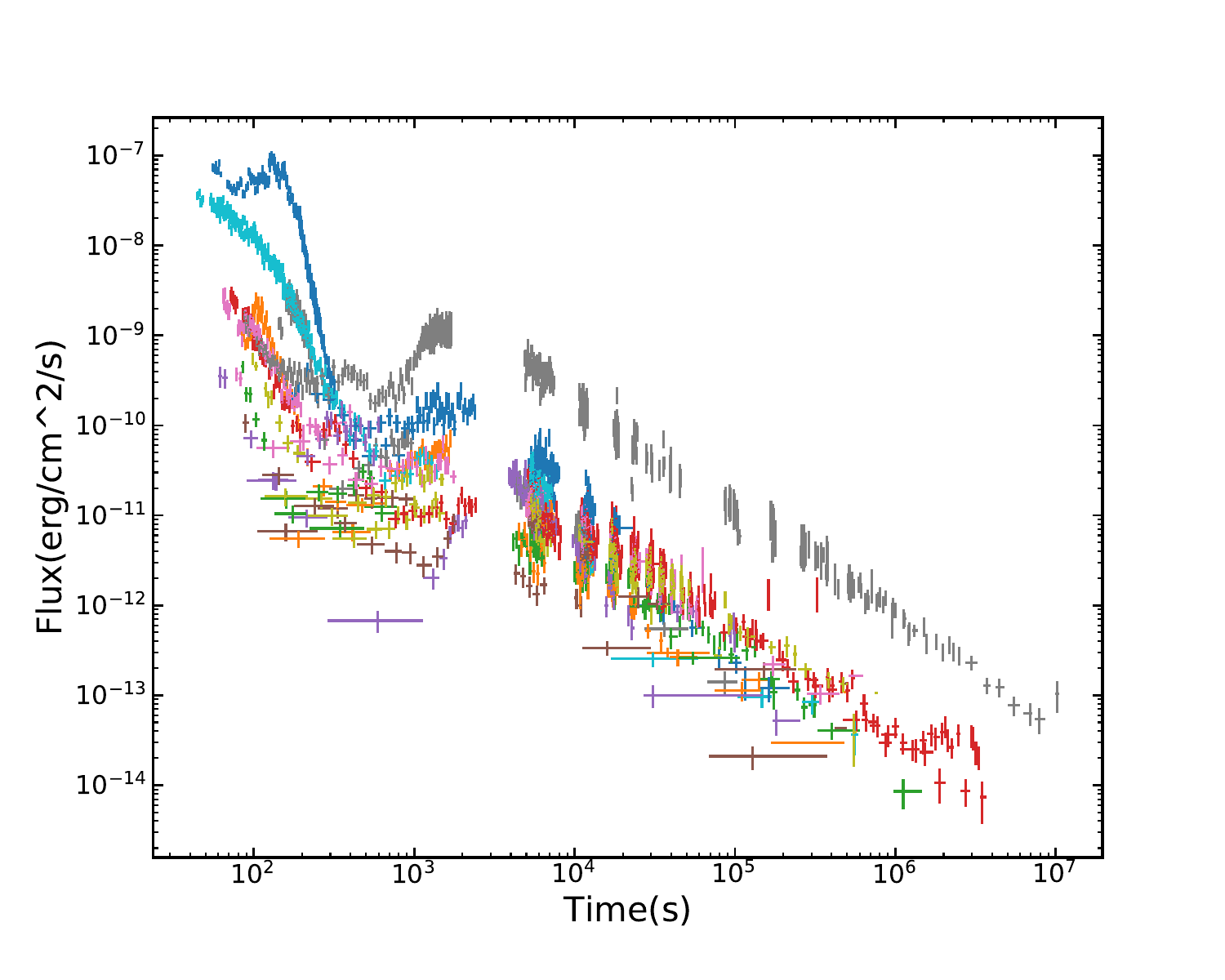}
    \includegraphics[width=3.5in]{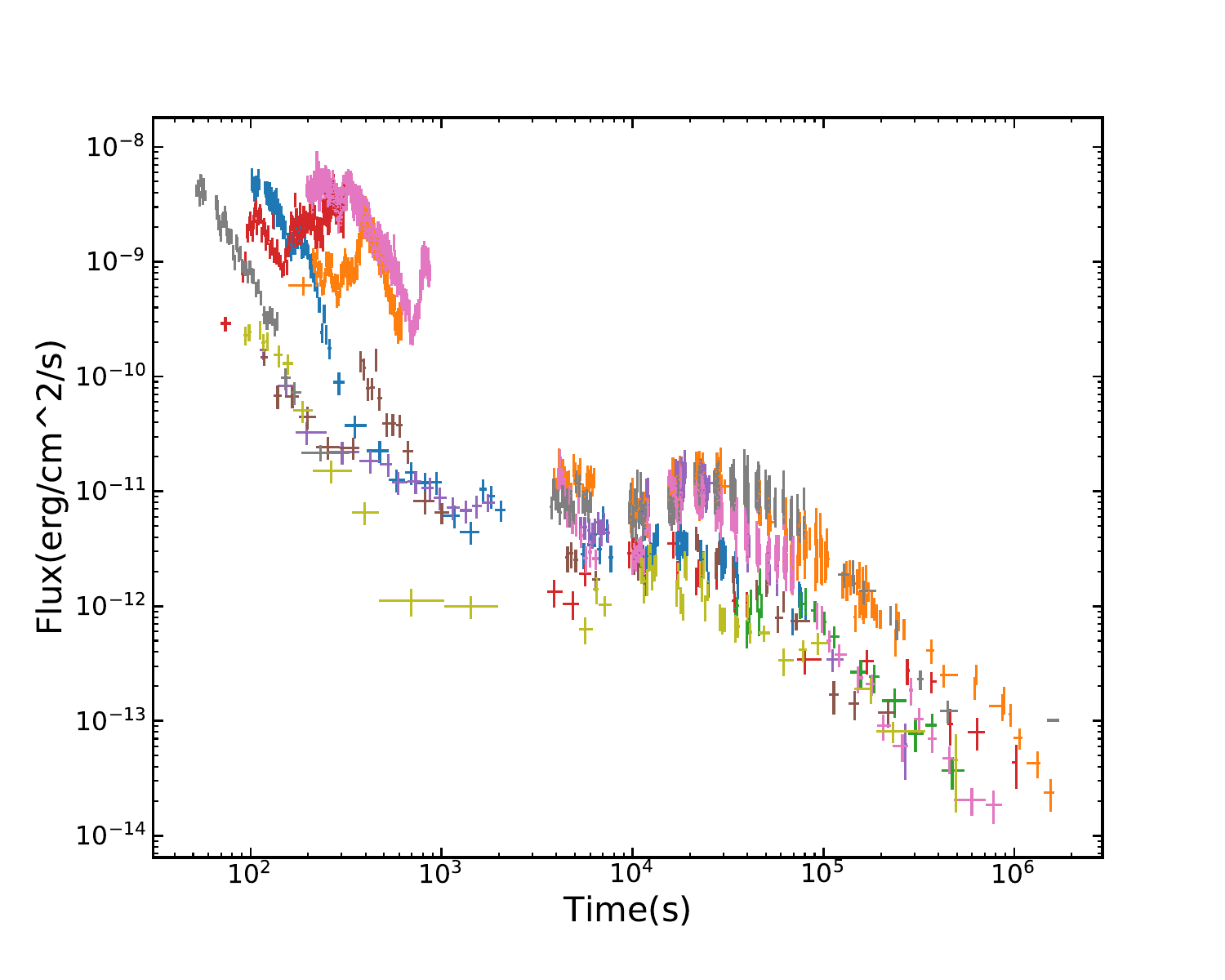}
    \caption{(a): X-ray light curves of our total sample. (b): Fits of our sample with the broken power-law model. (c): Histogram of the peak time distribution for our sample. The dashed lines are the Gaussian fits. (d): X-ray light curves of early bumps. (e): X-ray light curves of late bumps.}
        \label{Fig1}
\end{figure*}

In general, the X-ray emission of GRBs is very complicated \cite{2006ApJ...642..354Z}. Since, the steep decay segment with a power-law decay is from the curvature effect as the prompt emission tail. The light curves of X-ray are considered to fit by the initial steep decay component, afterglow component (e.g., from external shock) with power-law decay, smooth broken power-law segment (bumps), as well as the post-jet segment if it needs possible. Our sample do not include the cases in \citep{2016ApJS..224...20Y} who performed a systematic study and defined the X-ray flares. So that, excepting the GRBs that are in \citep{2016ApJS..224...20Y}, three criteria are adopted for our sample selection. (i) the rising and decay slopes of X-ray bumps are required to be less than that of X-ray flares defined in \cite{2016ApJS..224...20Y}. (ii) the duration of X-ray bumps should be longer and wider than that of X-ray flares in \cite{2016ApJS..224...20Y}. (iii) in order to extract the X-ray light curve, the observed X-ray light curve should be at least more than 5 data points. By accepting above criteria, our sample therefore only comprises 28 long-duration GRBs, including 17 GRB with redshift measured. Figure \ref{Fig1} shows two examples of the fitting results with smooth broken power-law function and other components\footnote{ https://astro.gxu.edu.cn/info/1062/2243.htm}. The fitting results are presented in Table \ref{tb1}, it includes the rising and decay slopes of bumps, peak times, and the start and end times of bumps.

Fitting the X-ray light curve with a broken power-law model, one finds that peak time of X-ray bumps exhibits bimodal distribution with peak times as $t_{\rm p,1}=1273\pm 686$ s and $t_{\rm p,2}=21752\pm 8566$ s, respectively. The division line is at $t=7190$ s (see Figure \ref{Fig1}). So that, we classify the X-ray bumps as two categories, e.g., “early” bumps with $t_{\rm p}<7190$ s, and “late” bumps with $t_{\rm p}>7190$ s, respectively. Moreover, one needs to clarify that the bimodal distribution of peak times can be affect by the selection effect, namely, the number of what we selected sample may be a sun-class of total GRBs observed by Swift/XRT. On the other hand, it is also possible affect by the sheltering from earth, namely, the dip of bimodal distribution of apparent peak times is possible caused by the selection effect due to sheltered time from earth.

\section{Models description of magnetar and black hole fall-back accretion}
In this section, we will present more details in theory for fall-back accretion onto a new-born
magnetar and black hole, respectively.
\subsection{Hyper-accretion fall-back disk around a newborn millisecond magnetar model}
A millisecond magnetar may survive as the central engine of long-duration GRBs after the massive star collapse. A small fraction of ejecta can not escape from the system due to the gravity of central magnetar, and it will fall back to the surface of magnetar. If this is the case, the central magnetar will spin up again when the accumulated materials on the surface of magnetar is large enough. The fall-back accretion onto a new-born magnetar is also used to interpret the early X-ray and optical bumps in the GRB afterglow \citep{1998PhRvL..81.4301D, 1998A&A...333L..87D, 2012ApJ...759...58D, 2015ApJ...805...88Y, 2016ApJ...831....5Z}. Roughly estimated, the minimum time scale of fall-back is equal to free-fall time scale. In fact, the time scale of fall-back is affected by many factors \citep{2001ApJ...550..410M,2012ApJ...759...58D}. In our calculations, we adopt $t_{0}\sim10^2$ s which is derived from numerical simulations as the starting time of materials fall-back \citep{2001ApJ...550..410M}. 

Following the method in \cite{2001ApJ...550..410M} and \cite{2008ApJ...683..329Z}, the acceleration rate of
fall-back can be expressed as
\begin{equation}
\dot{M}={({\dot{M}^{-1}}_{\rm early}+{\dot{M}^{-1}}_{\rm late})}^{-1}
\end{equation}
where
\begin{equation}
\dot{M}_{\rm early}=10^{-3}\eta_{\rm mag} t^{1/2}M_\odot~s^{-1}
\end{equation}
and
\begin{equation}
\dot{M}_{\rm late}=10^{-3}\eta_{\rm mag} {t_1}^{13/6}t^{-5/3}M_\odot~s^{-1}
\end{equation}
Here, $\eta_{\rm mag} \sim 0.01 - 10 $ is a factor that accounts for different explosion energies, and $t_1 \sim {\rm 200 - 1000 s}$ is similar to the time of peak flux of radiation generated by fall-back accretion (longer $t_1$ corresponding to smaller $\eta_{\rm mag}$). One roughly estimates $\dot{M}\propto t^{-5/3}$ if $t \gg t1$ \citep{1989ApJ...346..847C}. The gravitational mass ($M$) of the central magnetar can be obtained as \cite{2012ApJ...759...58D},
\begin{equation}
M=M_b(t) \left[1+\frac{3}{5} \frac{GM_b(t)}{R_s c^2}\right],
\end{equation}
and
\begin{equation}
M_b(t)=M_0+\int _{0}^{t}\dot{M}dt,
\end{equation}
Here, $R_s$ and $M_0$ are the radius and the initial baryonic mass of the magnetar, respectively, $M_b$ is the total baryonic mass of magnetar at time $t$.

Based on the results of \cite{2012ApJ...759...58D}, three radii are defined within the accretion disk, e.g., co-rotation radius ($r_c$), magnetospheric radius ($r_m$), and radius of light cylinder ($R_L$).
The $r_c$ is defined as following when the rotating angular velocity ($\Omega_s$) of the central engine is equal to Keplerian angular velocity ($\Omega_k$),
\begin{equation}
r_{\rm c}=\left(\frac{GM}{\Omega_s^2}\right)^{1/3}.
\end{equation}
The magnetospheric radius $r_m$ is
\begin{equation}
r_{\rm m}=\left(\frac{\mu^4}{GM\dot{M}^2}\right)^{1/7},
\end{equation}
where $\mu=B_0R_s^3$ is the magnetic dipole moment of the magnetar, and $B_0$ is the initial surface magnetic field of magnetar. The radius of the light cylinder is defined as,
\begin{equation}
R_{\rm L}=\frac{c}{\Omega_s}.
\end{equation}
For simple calculations, one define the fastness parameter as
\begin{equation}
\omega=\frac{\Omega_s}{\Omega_K(r_{\rm m})}=\left(\frac{r_{\rm m}}{r_{\rm c}}\right)^{3/2}.
\end{equation}
So that, the net torque exerted on the magnetar by the accretion disk reads as\footnote{Here, we do not consider the propellor effects of magnetar which is suggested by \citep{2011ApJ...736..108P}, and it required due a high magnetic filed of magnetar (e.g., $\sim 5\times 10^{14}$ G). More details can also refer to \citep{2024ApJ...962....6Y}.}
\begin{equation}
\tau_{\rm acc}=n(\epsilon,\omega)\frac{\mu^2}{r_{\rm m}^3}
\end{equation}
where $\epsilon=(r_{\rm m}/R_{\rm L})^{3/2}$, and $n(\epsilon,\omega)$ is the dimensionless torque parameter,
$$n(\epsilon,\omega)=
\begin{cases}
(2-2\epsilon+6\omega+3\epsilon^2\omega)/(9\omega),\text{~ $\omega \textless 1$,}\\
(2-2\epsilon+6\omega+3\epsilon^2\omega-9\omega^2)/(9\omega),\text{~$\omega \geq 1$.}
\end{cases}$$
The spin evolution is given by the following differential equation,
\begin{equation}
\frac{d(I\Omega_s)}{dt}=\tau_{\rm acc}+\tau_{\rm dip}
\end{equation}
where $I=0.35MR_s^2$ is the stellar moment of inertia, and $\tau_{\rm dip}$ is the torque due to magnetic dipole radiation and the inclination angle ($\chi$) between the magnetic axis and rotation axis,
\begin{equation}
\tau_{\rm dip}=-\frac{\mu^2\Omega_{s}^3\rm sin^2\chi}{6c^3}=-\frac{\mu^2\rm sin^2\chi}{6R_{\rm L}^3}.
\end{equation}
By solving above equations, one can get the luminosity of magnetic-dipole-radiation as a function
of time \citep{1983bhwd.book.....S},
\begin{eqnarray}
L_{\rm dip}&=&\frac{\mu^2\Omega_s^4\rm sin^2\chi}{6c^3} \nonumber \\
&=&9.6\times10^{48}\rm erg\ s^{-1}sin^2\chi \left(\frac{\mu}{10^{33}\rm G~cm^3}\right)^2\left(\frac{P_0}{1\
\rm
ms}\right)^{-4}
\end{eqnarray}
Here, $P_0$ is the initial period of magnetar.

\subsection{Black hole fall-back accretion model}
Alternatively, a black hole may be formed as the central engine of long-duration GRBs after the massive star core collapse. The GRB jet can be powered by annihilating between neutrinos and anti-neutrinos that can carry the accretion energy in the disk, or extracts the spin energy of the black hole which can be tapped by a magnetic field connecting the outer world through the Blandfordd-Znajek (BZ) mechanism \citep{1977MNRAS.179..433B}. The fall-back accretion into black hole is also adopted to interpret the late X-ray and optical bumps in the GRB afterglow \citep{2013ApJ...767L..36W,2017ApJ...849...47L,2017ApJ...849..119C,2021ApJ...906...60Z}. 

The time scale of fall-back is approximately equal to the free-fall time scale, $t_{fb}\sim(\pi^2r_{fb}^3/8GM_\bullet)^{1/2}$, where $M_\bullet$ is the mass of black hole, $r_{\rm fb}$ is the radius of progenitor. The evolution of the fall-back accretion rate can be described by a broken power-law function of time \citep{1989ApJ...346..847C,2001ApJ...550..410M},
\begin{equation}
\dot{M} = \dot{M}_p \left[\frac{1}{2}\left(\frac{t-t_0}{T_p-t_0}\right)^{-s/2}+\frac{1}{2}
\left(\frac{t-t_0}{T_p-t_0}\right)^{5s/3}\right]^{-\frac{1}{s}}
\end{equation}
where $t_0$ is the beginning time of the fall-back accretion, and $s$ describes the sharpness of the peak. $T_p$ and $\dot{M}_p$ are the peak time and peak rate of fall-back accretion, respectively.

Based on the results of references \citep{2000PhR...325...83L,2000PhRvD..61h4016L,2002MNRAS.335..655W,2005ApJ...630L...5M,2011ApJ...740L..27L,2017ApJ...849..119C, 2017NewAR..79....1L}, the BZ power from a kerr black hole can be estimated as,
\begin{equation}
\dot{E}_{\rm BZ}=L_{\rm BZ}=1.7\times10^{50}{a_\bullet}^2{M_\bullet}^2{B_{\bullet,15}}^2F({a_\bullet})\rm
\enspace erg \enspace
s^{-1},
\end{equation}
where $M_\bullet$ is the mass of black hole with unit of $M_{\odot}$, $
B_{\bullet,15}=B_\bullet/10^{15}G$ is the magnetic filed, $a_\bullet=J_\bullet c/(G{M_\bullet}^2)$
and $J_\bullet$ are the spin and angular momentum of black hole, respectively.
\begin{equation}
F(a_\bullet)=[(1+q^2)/q^2][(q+1/q)\arctan q-1].
\end{equation}
Here $q=a_\bullet/(1+\sqrt{1-a_\bullet^2})$. The range of spin parameter is $0 \le a_\bullet \le
1$, so that, one has $2/3 \le F(a_\bullet) \le \pi-2$. By assuming that the magnetic field pressure of the BH and the ram pressure ($P_{\rm ram}$) of the innermost parts of an accretion flow are in
balance, one has \citep{1997rja..proc..110M}
\begin{equation}
\frac{B_\bullet ^2}{8\pi}=P_{\rm ram}\sim pc^2 \sim \frac{\dot{M}c}{4\pi r_\bullet ^2},
\end{equation}
where $r_\bullet = (1+\sqrt{1-a_\bullet^2})r_g$ is the radius of the BH horizon and
$r_g=GM_\bullet/c^2$. Combining with Equations (17), (18), and (19), $B_\bullet$ and $L_{\rm BZ}$
can be rewritten as
\begin{equation}
B_\bullet \approx 7.4\times10^{16}\dot{M}^{1/2}M_\bullet^{-1} (1+\sqrt{1-a_\bullet^2})^{-1}\rm G.
\end{equation}
\begin{equation}
L_{\rm BZ}=9.3\times10^{53}a_\bullet^2\dot{M}X(a_\bullet)\rm \enspace erg \enspace s^{-1},
\end{equation}
where
\begin{equation}
X(a_\bullet)=F(a_\bullet)/(1+\sqrt{1-a_\bullet^2})^2.
\end{equation}

Within the scenario of the energy conservation and angular momentum conservation, there are two teams of the evolution equation of BH in the BZ model, i.g., spin-up by accretion and spin-down by the BZ mechanism \citep{2002MNRAS.335..655W},
\begin{equation}
\frac{d M_\bullet c^2}{d t}=\dot{M}c^2E_{\rm ms}-L_{\rm BZ}
\end{equation}
\begin{equation}
\frac{d J_\bullet}{d t}= \dot{M}L_{\rm ms}-T_{\rm BZ}
\end{equation}
\begin{equation}
\frac{d a_\bullet}{d t}= (\dot{M}L_{\rm ms}-T_{BZ})c/(G{M_\bullet}^2)-2a_\bullet(\dot{M}c^2E_{\rm ms}-L_{\rm
BZ})/(M_\bullet c^2)
\end{equation}
where $T_{\rm BZ}$ is the total magnetic torque that applied to BH \citep{2000PhRvD..61h4016L,2011ApJ...740L..27L,2017ApJ...849...47L},
\begin{equation}
T_{\rm BZ}=\frac{4GM_{\bullet}\cdot L_{\rm BZ}\cdot(1+\sqrt{1-a_{\bullet}^2})}{a_{\bullet}c^{3}}.
\end{equation}
In equation (23) and (24), $E_{\rm ms}$ and $L_{\rm ms}$ are the specific energy and angular
momentum which are corresponding to the innermost radius ($r_{\rm ms}$) of the disk, and defined as
following,
\begin{equation}
E_{\rm ms}=\frac{4\sqrt{R_{\rm ms}}-3a_\bullet}{\sqrt{3}R_{\rm ms}}
\end{equation}
\begin{equation}
L_{\rm ms}=\frac{GM_\bullet}{c}\frac{(6\sqrt{R_{\rm ms}}-4a_\bullet)}{\sqrt{3R_{\rm ms}}}
\end{equation}
and $R_{\rm ms}$ can be referenced from \cite{1972ApJ...178..347B},
\begin{equation}
R_{\rm ms}=\frac{r_{\rm ms}}{r_{\rm g}}=3+Z_2-[(3-Z_1)(3+Z_1+2Z_2)]^{1/2}
\end{equation}
Here, $Z_1\equiv 1+(1-a_\bullet^2)^{1/3}[(1+a_\bullet)^{1/3}+(1-a_\bullet)^{1/3}]$ and $Z_2 \equiv
(3a_\bullet^2+Z_1 ^2 )^{1/2}$  \enspace($0\le a_\bullet \le 1$).
\section{The fitting results of models application to our selected sample}
In this section, by assume that the early and late bumps of X-ray afterglow are related to the fall-back accretion of magnetar and black hole, respectively. Then, we adopt the fall-back accretion of magnetar and black hole to fit the early and late bumps of X-ray afterglow with Monte
Carlo Markov Chain (MCMC) method, respectively. Due to no redshift measured of some GRBs in our sample, we adopt $z=1$ to do the calculations instead of no redshift-measured GRBs\footnote{\cite{2006A&A...447..897J} found that the mean redshift is $z\sim2.8$ for a small GRB samples. The different pseudo-redshift what we adopt is indeed making a huge difference of inferred energetics, and it can affect the inferred parameters of fall-back accretion in both magnetar and black hole}. 

\subsection{Fall-back accretion of magnetar to early bumps}
Based on the criteria of sample selected and the distribution of peak time for both early and late X-ray bumps, there are 19 long-duration GRBs which are identified as “early bump” category, and 12 of which have redshift measurements. So that, we attempt the fall-back accretion of magnetar model to fit the those GRBs with early bumps. Considering the poorly understanding the equation of state of magnetar, we adopt the typical values of mass and radius of magnetar as 1.4$M_\odot$ and 10 kilometers to do the fits, respectively. Moreover, we also fix $\rm sin\chi=0.5$ in our calculations due to unknown inclination angle of the magnetic axis to the rotation axis \citep{2012ApJ...759...58D}.

\begin{figure*}
    \includegraphics[width=3.5in]{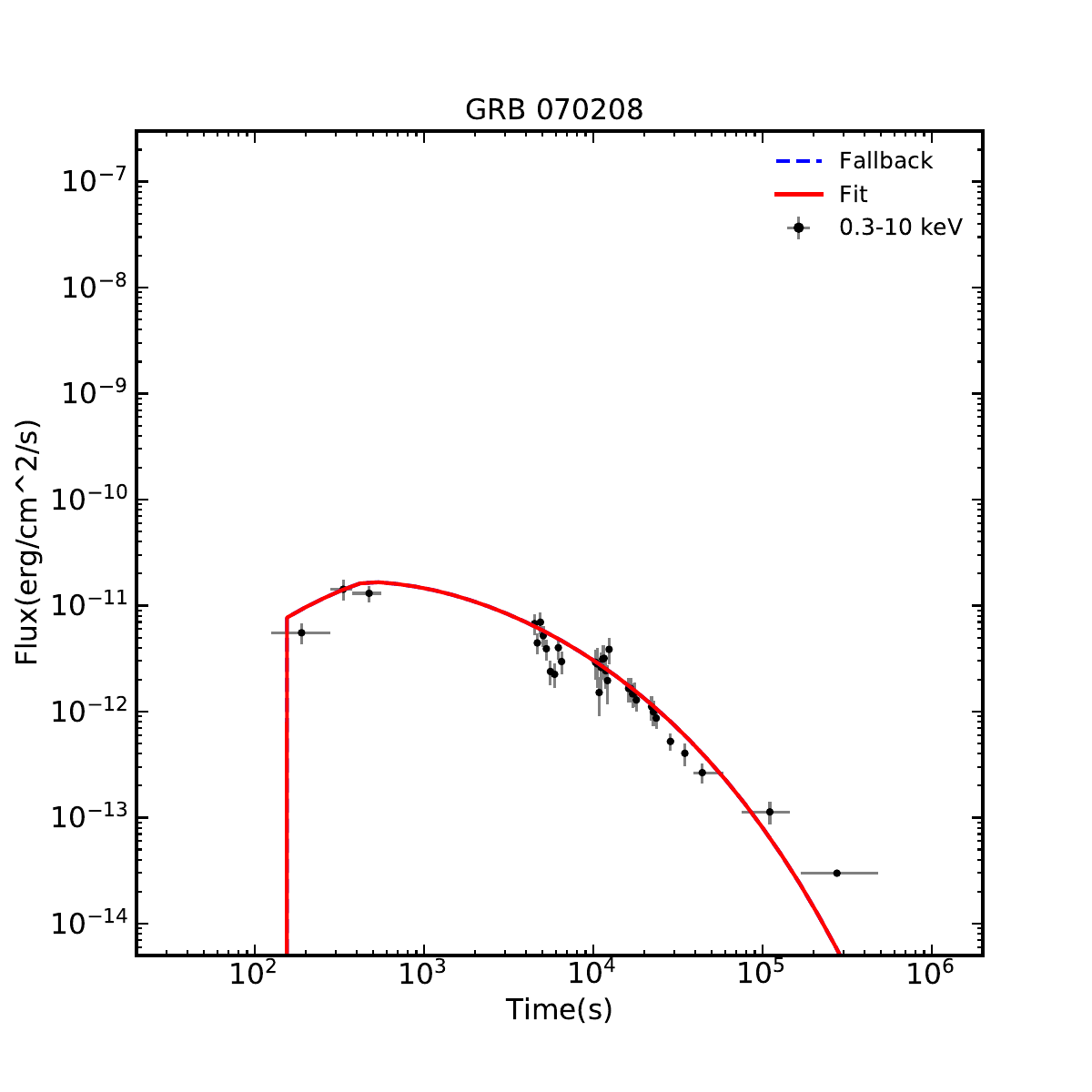}
    \includegraphics[width=3.5in]{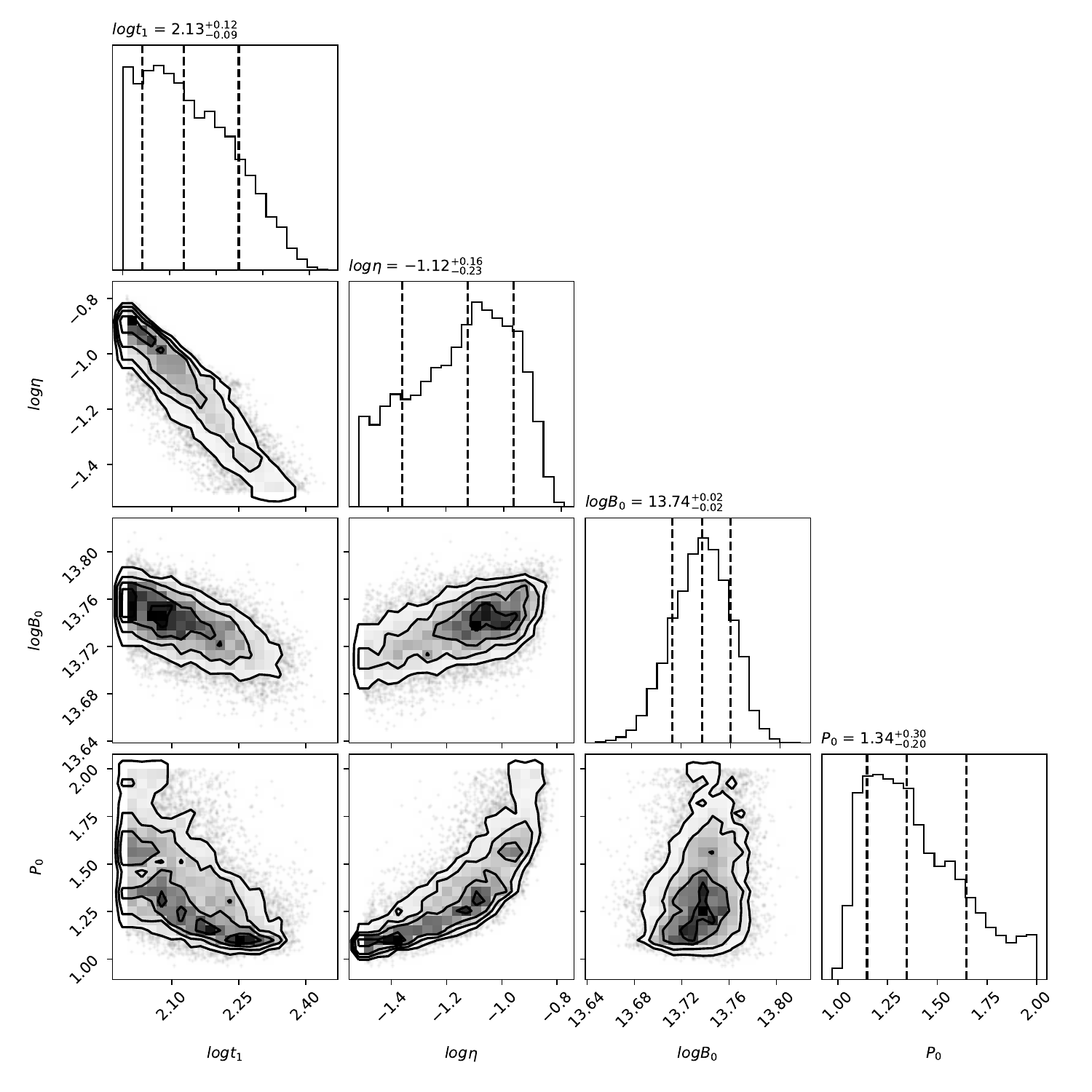}
    \includegraphics[width=3.5in]{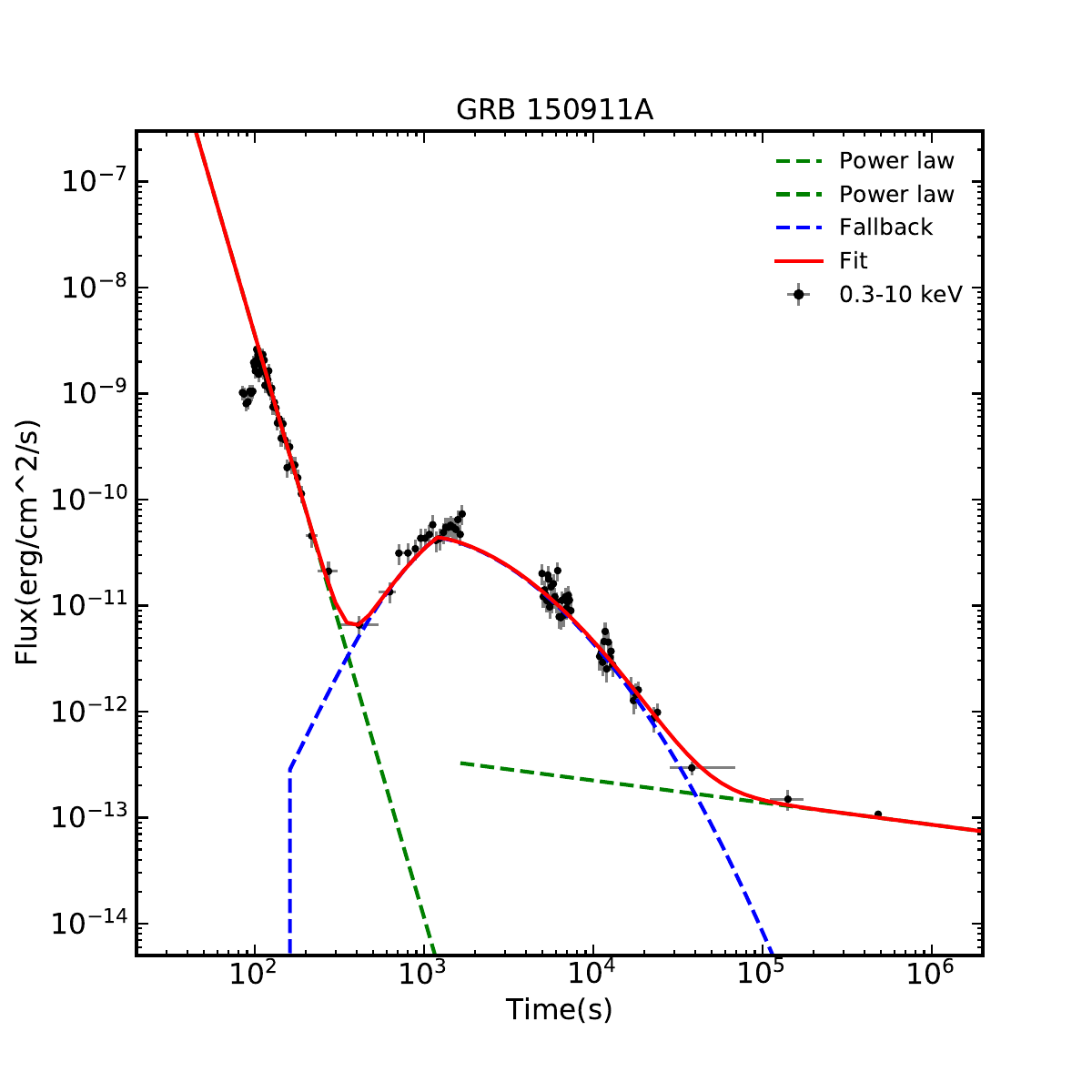}
    \includegraphics[width=3.5in]{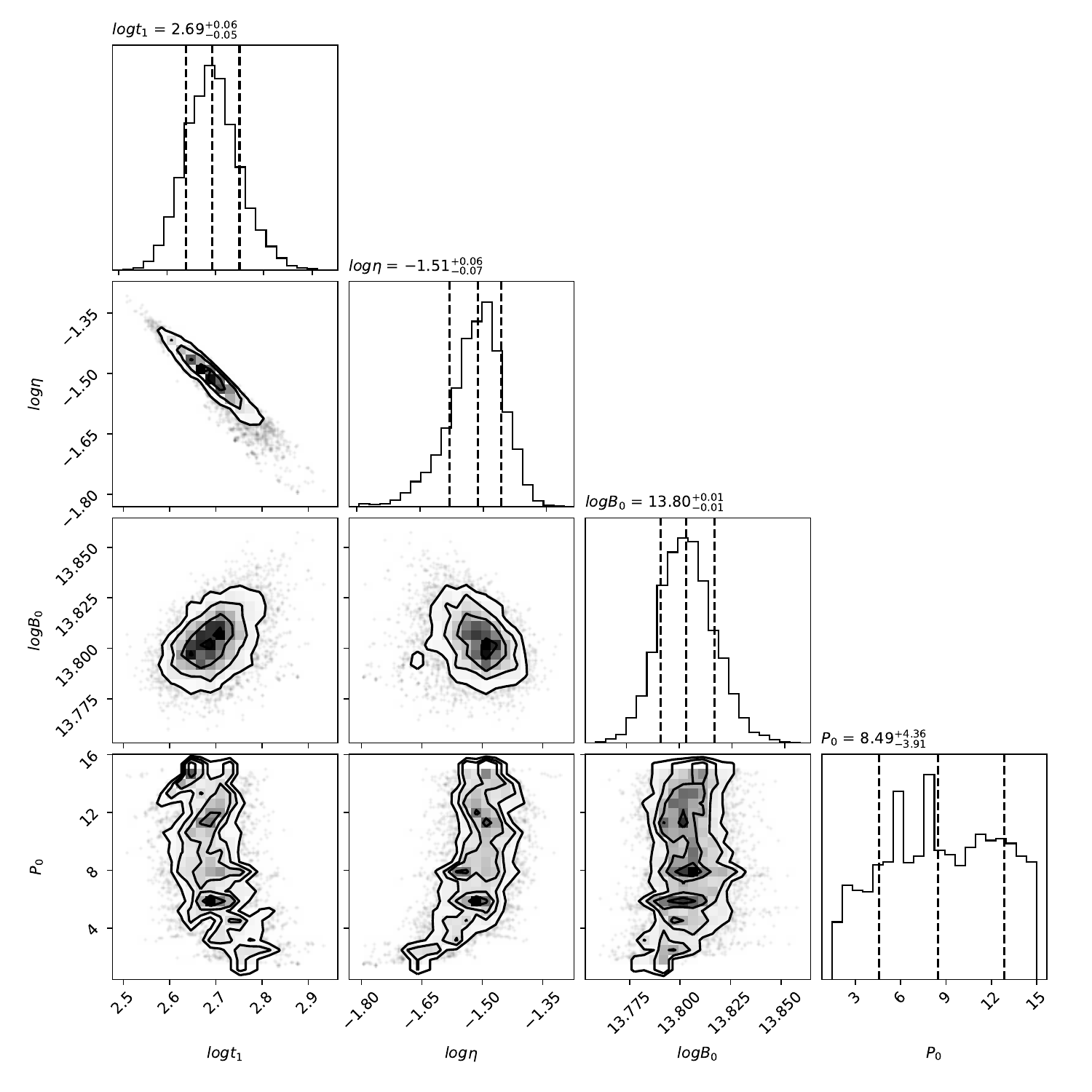}
    \caption{Fall-back accretion of magnetar modeling results of two examples for the early bumps of X-ray afterglow (left) and the corner plots of the free parameters posterior probability distribution.}
        \label{Fig2}
\end{figure*}

If this is the case, there are four free parameters of fall-back accretion of magnetar, e.g., $t_1$, $\eta_{\rm mag}$, $B_0$, and $P_0$. Moreover, by considering the contributions from the initial steep decay segment (tail emission of prompt emission) and afterglow with power-law decay component (e.g., external shock) to the fits, we also simultaneously fit the light curves with all three components above (e.g., steep decay, afterglow, and fall-back accretion). Then, we adopt the MCMC method with {\em Python code} to fit the data. In our fitting, we use a {\em Python module emcee} to get best-fit values and uncertainties of free parameters \citep{2013PASP..125..306F}. The allowed range of the four free parameters are set as $\rm log(t_1)\epsilon [2.15, 3.36]$ s, $\rm log(\eta_{\rm mag})\epsilon [-2.16, -1.05]$, $\rm log(B_0)\epsilon[13.58,13.96]$; and $P_0\epsilon[0.8,8.5]$ ms. Figure~\ref{Fig2} shows two examples of best fitted light curves and the corner plots of free parameters posterior probability distribution for the fall-back accretion model\footnote{https://astro.gxu.edu.cn/info/1062/2244.htm}. The results of fits for all 21 early bumps are listed in Table \ref{tb2}. 

\begin{figure*}
    \includegraphics[width=3.5in]{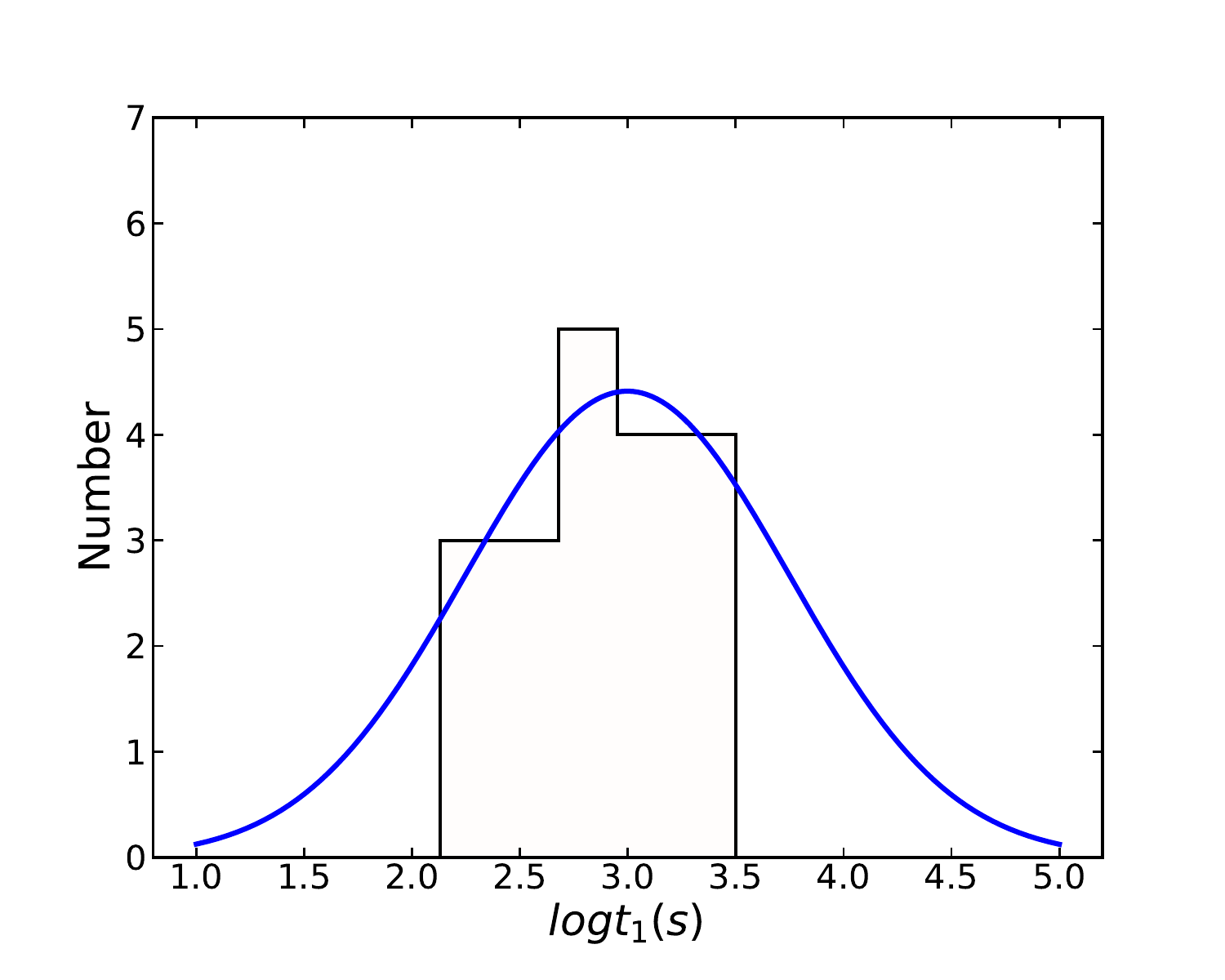}
    \includegraphics[width=3.5in]{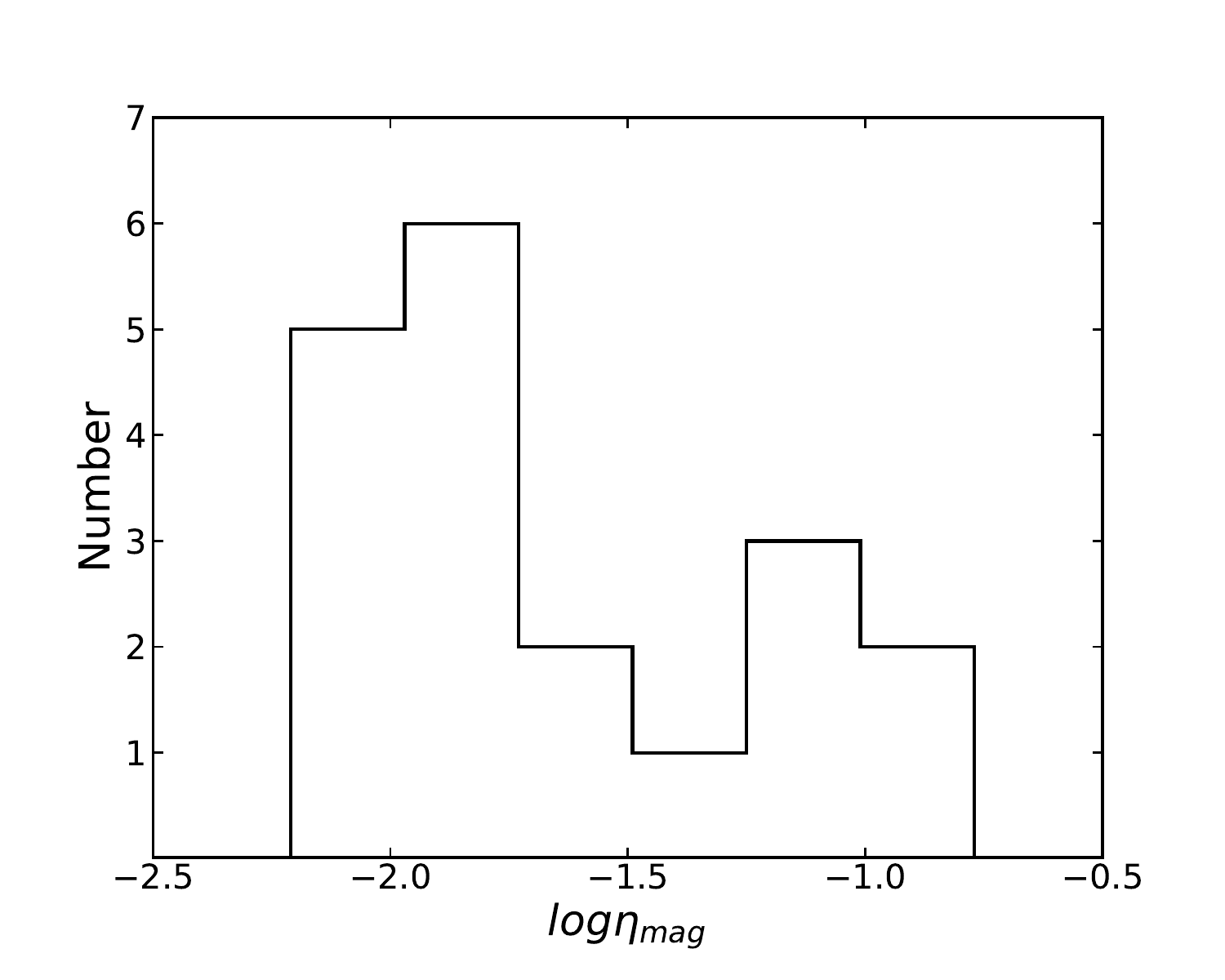}
    \includegraphics[width=3.5in]{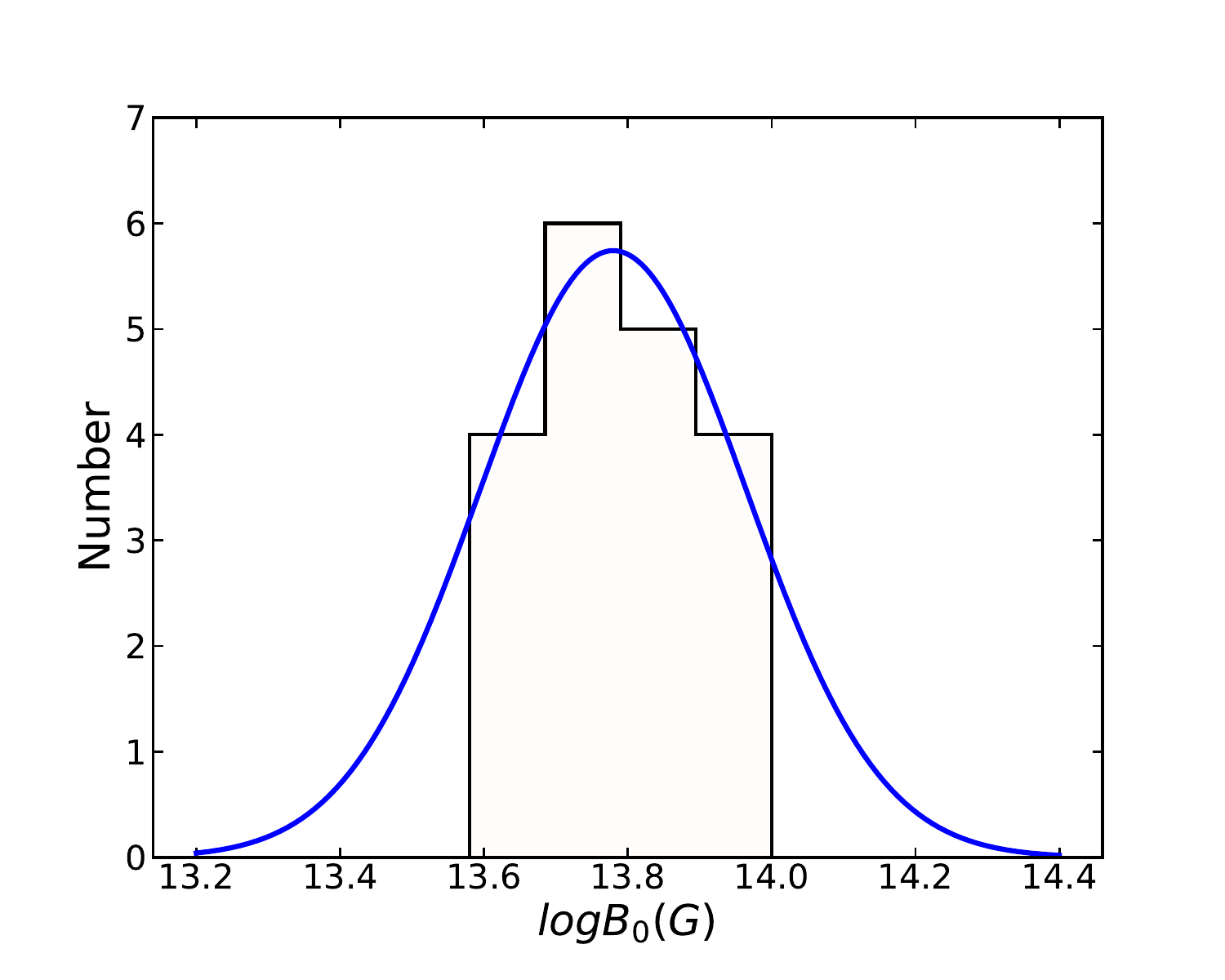}
    \includegraphics[width=3.5in]{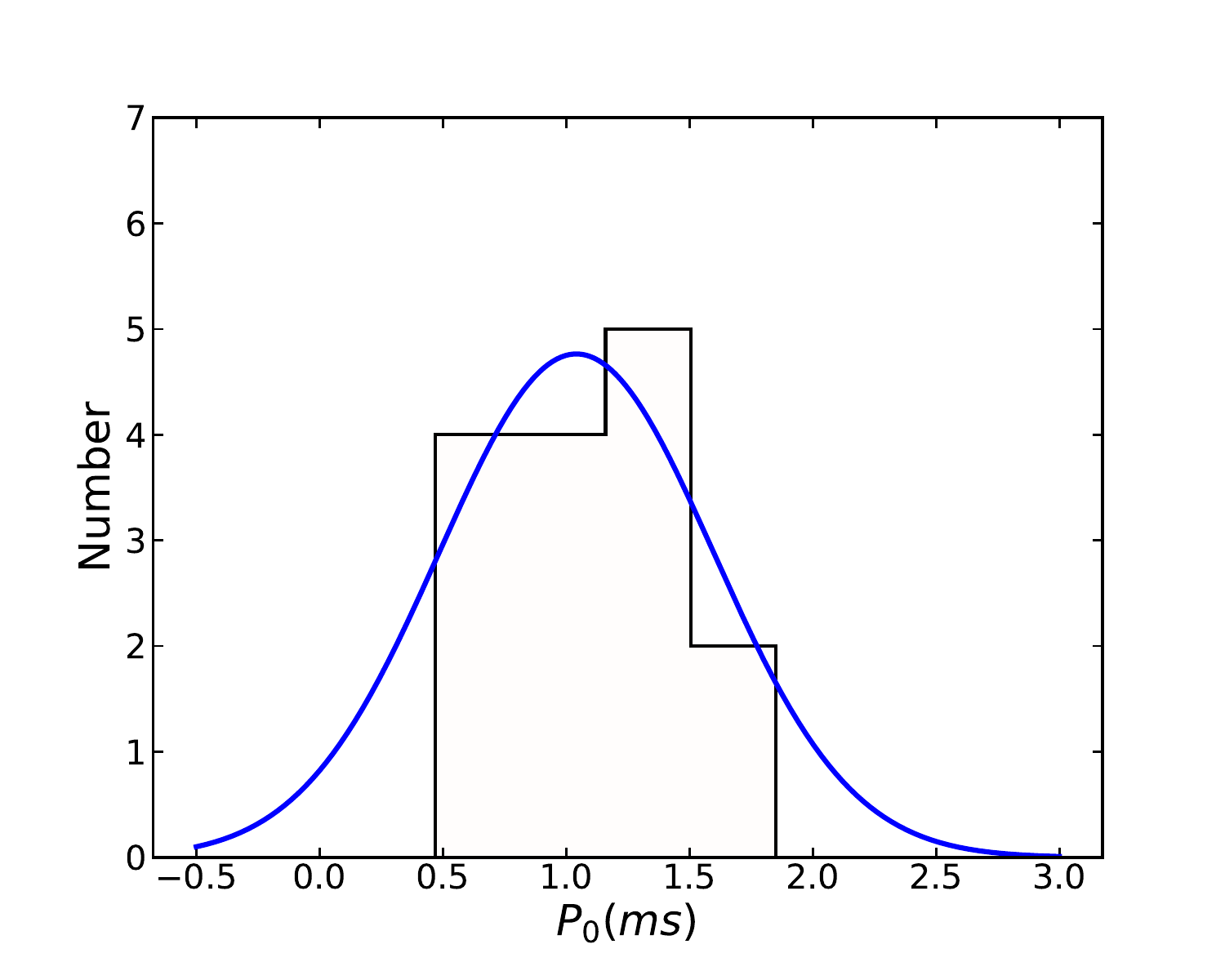}
    \caption{Distributions of free parameters for early bumps with fall-back accretion of magnetar model. The solid bule lines are the Gaussian fits.}
        \label{Fig3}
\end{figure*}

It is worth testing that what are the distributions of the values for $t_1$, $B_0$, and $P_0$. Figure~\ref{Fig3} shows the distributions of those fitting values. It is found that those values are obeyed the normal ($P_0$) or log-normal ($t_1$ and $B_0$) distributions with $\rm log (t_1)=(3.01\pm0.12)$ s, $\rm log (B_0)=(13.77\pm0.03)$ G, and $P_0=(1.04\pm0.05)$ ms, respectively. Moreover, the distribution of $\eta_{\rm mag}$ is range of [0.6\%, 17\%]. In any case, based on the MCMC fitting results, one can see that the early bumps of our sample are well fitted by fall-back accretion of magnetar model, and all free parameters can be constrained very well. The values of those parameters also fall into a reasonable range.

\subsection{Fall-back accretion of black hole to late bumps}
Based on the bimodal distribution of peak time of bumps in our sample, there are 9 GRBs which are identified as “late bump” category, and 5 of which have redshift measurements. So that, we apply the fall-back accretion of black hole model and the afterglow component to fit those GRBs with late bumps. Here we adopt the beginning time ($\rm t_s$) and ending time of the bumps ($\rm t_e$) as the beginning time and ending time of the fall-back accretion \citep{2021ApJ...906...60Z}. The initial mass and spin of black hole are fixed as $M_0 =3M_\odot$ and $a_0 = 0.9$, respectively \citep{2017ApJ...849...47L}. Meanwhile, due to uncertainty efficiency and jet opening angle of GRBs, we take $\eta_{\rm BH}= 0.01$ and $f_b = 0.01$ during the calculations. We define a dimensionless parameter as peak of fall-back accretion rate $\dot{m}_p$ ($\dot{m}_p=\dot{M}_p/M_\odot \rm \enspace s ^{-1}$), and $T_p$ is the peak time when the fall-back accretion rate $\dot{m}_p$ reaches at the peak. Here, we only focus on free parameters of fall-back accretion of black hole, e.g., $\dot{m}_p$, $T_p$, and $s$.

\begin{figure*}
    \includegraphics[width=3.5in]{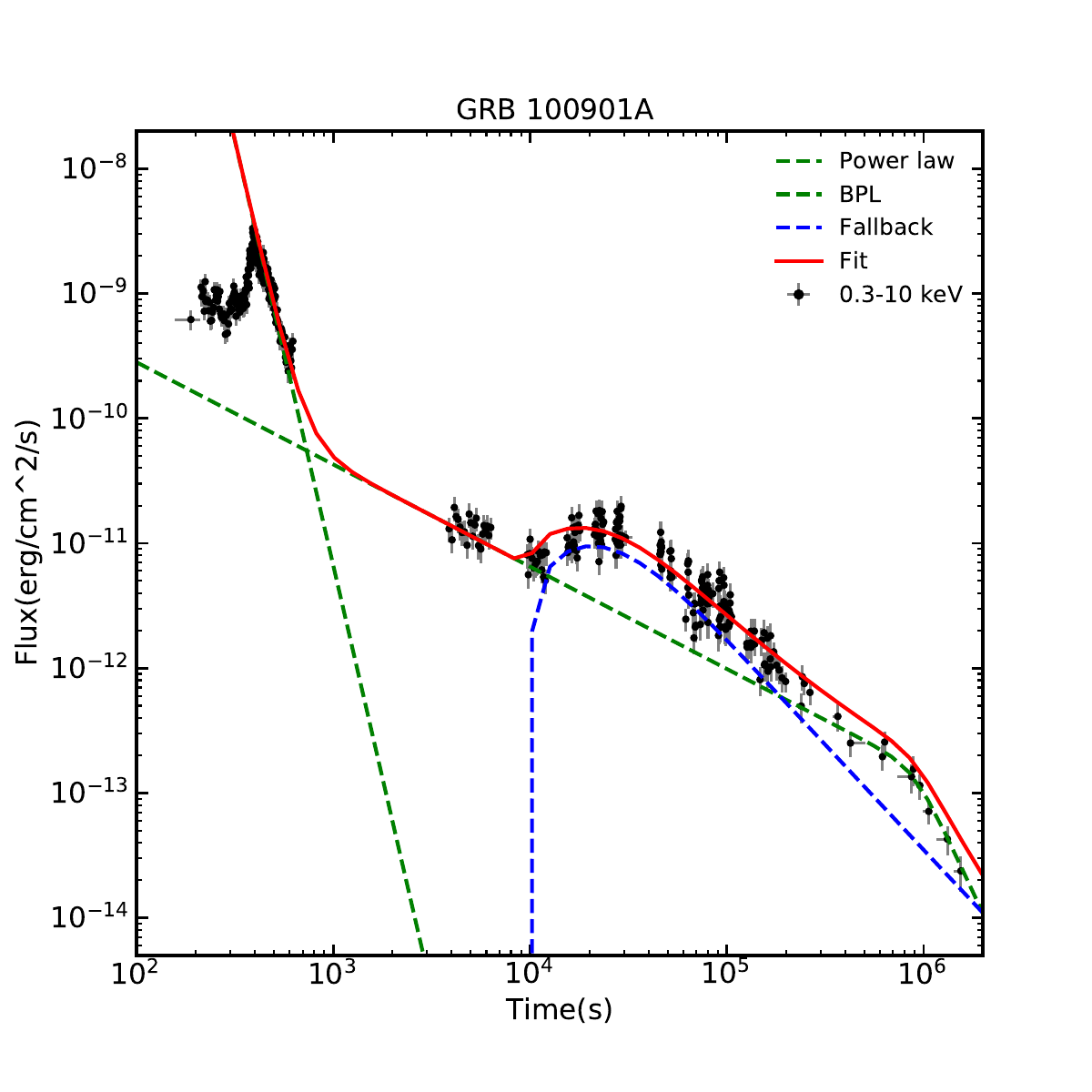}
    \includegraphics[width=3.5in]{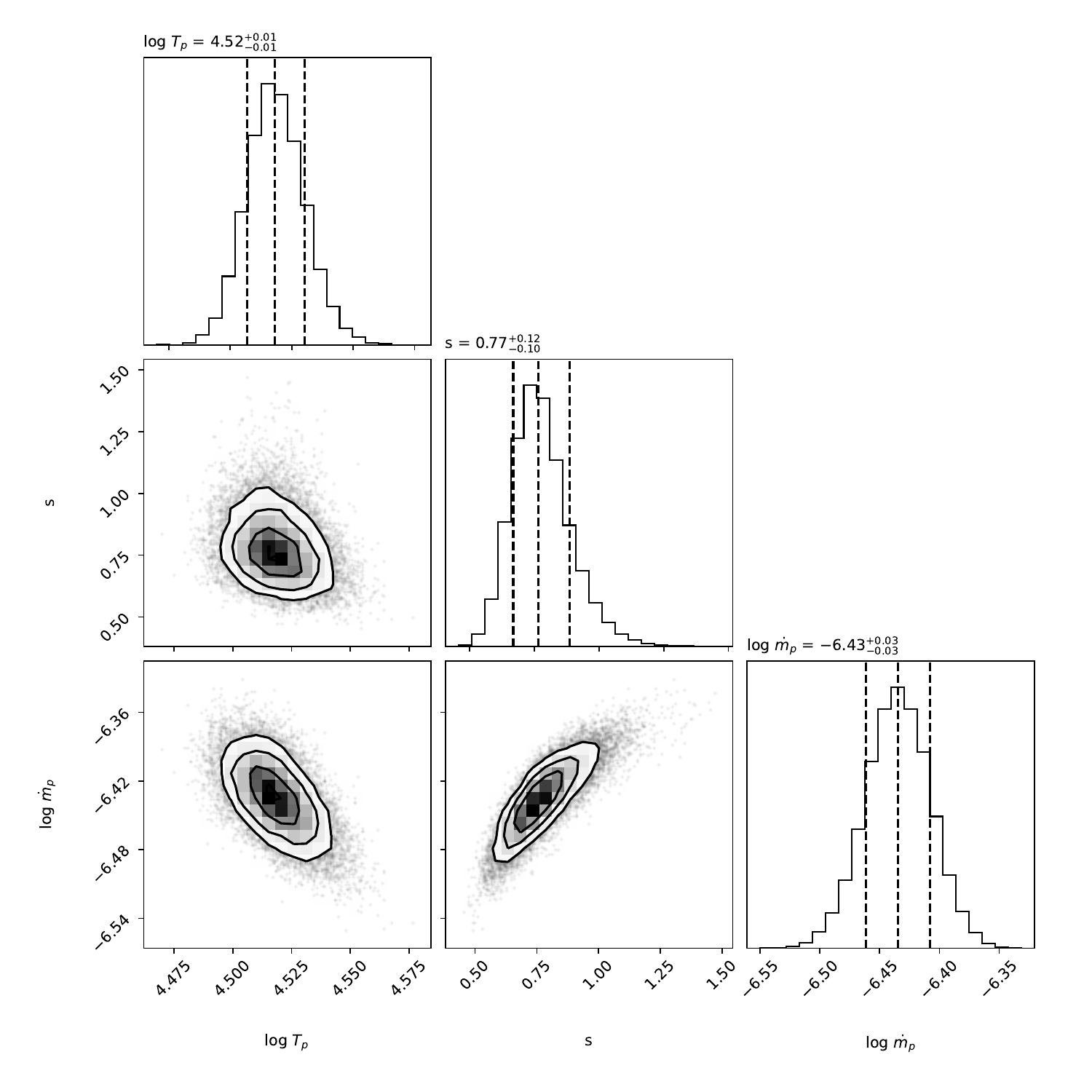}
    \includegraphics[width=3.5in]{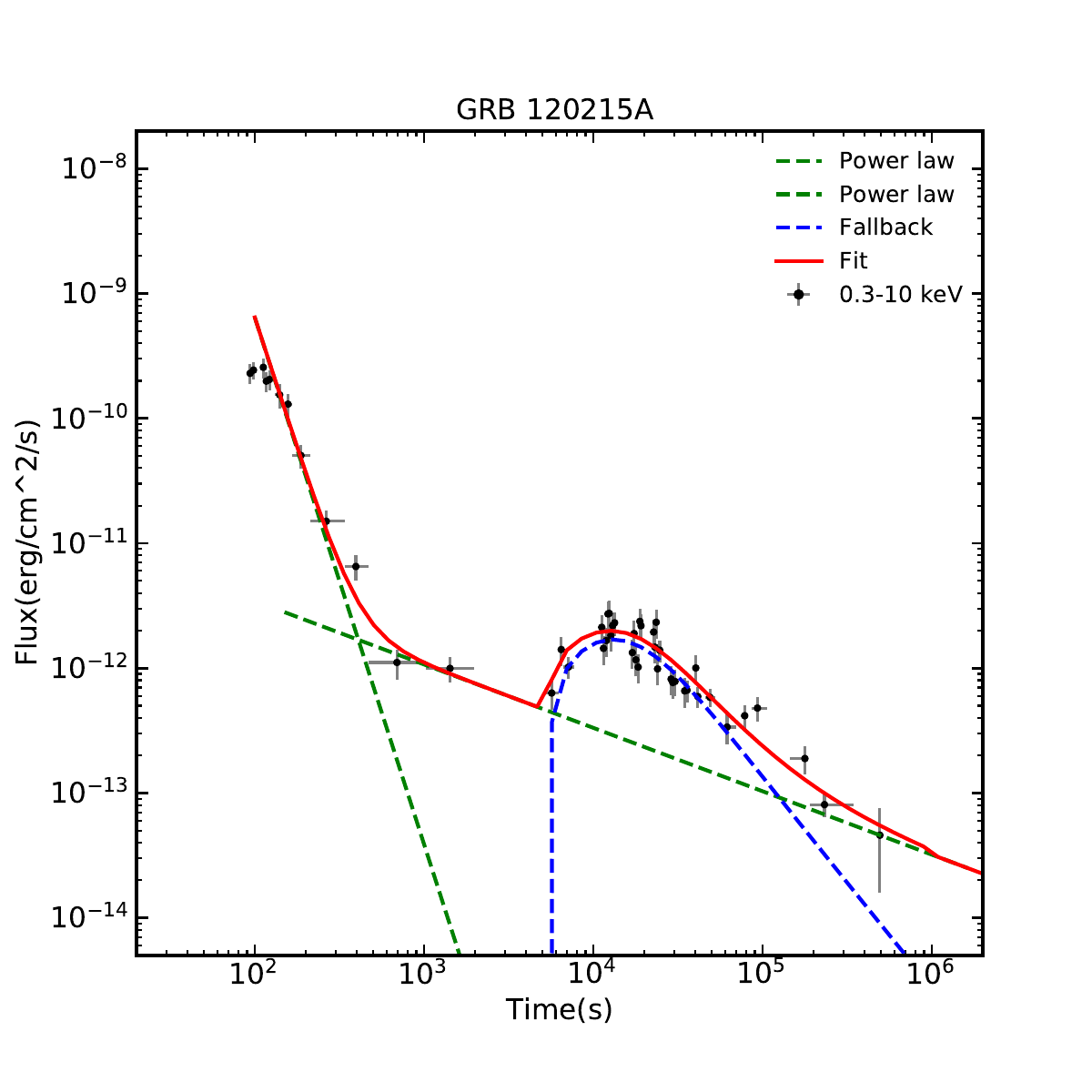}
    \includegraphics[width=3.5in]{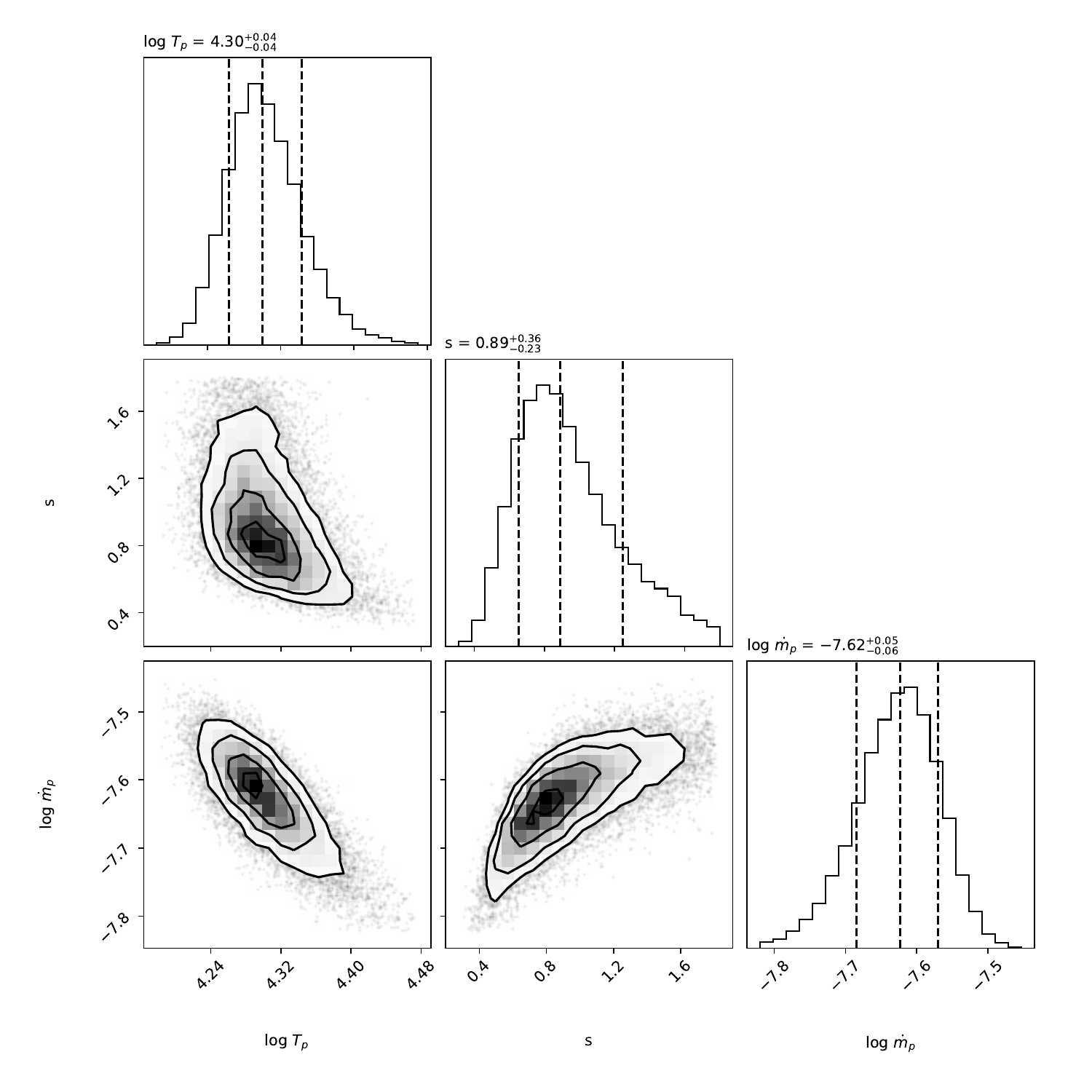}
    \caption{Fall-back accretion of black hole modeling results of two examples for the late bumps of X-ray afterglow (left) and the corner plots of the free parameters posterior probability distribution.}
        \label{Fig4}
\end{figure*}

We then adopt the MCMC method with {\em Python code} to fit the data of those 9 GRBs with late bumps. The ranges of the free parameters are set as follows: log$ (\dot{m}_p)\epsilon[-10,0]$, $s\epsilon[0,5]$ and $\rm T_p\epsilon[t_{s},t_{e}]$. Two examples of best-fitted light curves for the late-bumps and the the corner plots of free parameters posterior probability distribution are shown in Figure~\ref{Fig4}, and the full fitting results can be found in footnote 5. The results of fits for all 9 late bumps are listed in Table \ref{tb3}. It is found that the the distributions of $log~T_{\rm p}$, $s$, and $\rm log~\dot{m}_p$ are range of [4.26, 4.88], [0.33, 1.53], and [-7.77, -5.73], respectively.

Furthermore, we also calculate the total mass of the accretion ($M_{\rm acc}$) in the fall-back process when we do the integral of time for equation (15), and the fall-back radius can also be estimated as,
\begin{equation}
R_{\rm fb}\approx3.5\times10^{10}(M_\bullet/3M_\odot)^{1/3}(t_{fb}/360\  s)^{2/3}\rm cm
\end{equation}
The derived physical parameters of $M_{\rm acc}$ and $R_{\rm fb}$ are also listed in Table \ref{tb3}. We also do the distributions of the derived physical parameters, and it is found that the distribution of $M_{\rm acc}$ is range of $\rm [4\times 10^{-4}, 1.8\times 10^{-2}]~M_\odot$. The fall-back radius is range of $[1.04, 4.23]\times 10^{11}$ cm, which is consistent with the typical radius of a Wolf-Rayet star. In any case, the late bumps of our sample can be well fitted by fall-back accretion of black hole model, and the values of those parameters fall into a reasonable range.

Moreover, it is notice that both accretion rate and accretion mass in this work are much lower than that of in supernova explosion at the same time \citep{2019ApJ...880...21M}. That is because the adopted mass of progenitor star and black hole in \cite{2019ApJ...880...21M} are much larger than that of what we adopt.

\section{Conclusion and Discussion}
The central engine of long-duration GRBs remain an open question \citep{2011CRPhy..12..206Z}, and some characteristic of afterglow emission (i.e., X-ray re-brightening) may take a clue to understand the naturally central engine and progenitor of GRBs. In this paper, we systemically search for long-duration GRBs that consist of bumps in X-ray afterglow detected by Swift/XRT between 2005 January and 2023 July, and found that 28 candidate GRBs showing X-ray bumps in their afterglow. 

More interestingly, we find that the peak time of the X-ray bumps in our sample exhibit bimodal distribution, and defined as “early” and “late” bumps with division line at $t=7190$ s, e.g., early bumps with $t_{\rm p}<7190$ s, and late bumps with $t_{\rm p}>7190$ s, respectively. We proposed that the long-duration GRBs with an early (or late) bumps may be originated from the fall-back accretion onto a new-born magnetar (or black hole). By adopting MCMC method to fit the early (or late) bumps of X-ray afterglow with the fall-back accretion of magnetar (or black hole), we are able to reach several interesting results.
\begin{itemize}
 \item Both early and late bumps of X-ray afterglow in our sample can be well fitted by the fall-back accretion of magnetar and black hole, respectively. The values of parameters for those two models also fall into a reasonable range.
 \item The initial surface magnetic filed and period of magnetars for most early bumps are clustered around $5.88\times10^{13}$ G and $1.04$ ms, respectively.
 \item The fall-back accretion rate reaches its peak value $~(1.7\times 10^{-8}-1.9\times10^{-6})~M_{\odot}\rm ~s^{-1}$, and the derived accretion mass of black hole for late bumps is range of $[4\times10^{-4}, 1.8\times10^{-2}]~M_{\odot}$. The typical fall-back radius is distributed range of $[1.04, 4.23]\times 10^{11}$ cm which is consistent with the typical radius of a Wolf-Rayet star. 
\end{itemize}

A very interesting question is whether the early bumps can also be interpreted by fall-back accretion of black hole. In order to test this possibility, we invoke the same MCMC method to fit the early bumps of X-ray afterglow in our sample by using the fall-back of black hole central engine model. We find that most early bumps (15 out of 19) can be fitted by the fall-back accretion of black hole model, but require a larger accretion rate. It means that the fall-back accretion of black hole model can not be ruled out to interpret the early bumps. On the contrary, we also invoke the fall-back accretion magnetar model to fit the late bumps of X-ray afterglow in our sample. We find that the fall-back accretion magnetar model is very difficult to fit the late bumps, such as the parameters of model are not convergent, or unreasonable distribution of model parameters. It suggest that the fall-back accretion magnetar model is disfavored by the late bumps of X-ray afterglow in our sample.

In addition, several proposed models are also invoke to interpret the re-brightening feature in afterglow of GRBs, e.g., the fireball decelerated by the ambient medium \citep{1999ApJ...517L.109S,2007ApJ...655..973K}, the density bumps or voids in the circumburst medium \citep{2002ApJ...565L..87D,2002A&A...396L...5L}, a refreshed shock \citep{2002ApJ...566..712Z,2004ApJ...615L..77B}, a structured jet with off-axis \citep{2003JCAP...10..005N,2003Natur.426..154B,2004ApJ...605..300H,2008MNRAS.387..497P,2009A&A...499..439G,2010MNRAS.406.2149M}, a long-lasting reverse shock model \citep{2012ApJ...761..147U}, the existence two component jets \citep{2004ApJ...605..300H,2005MNRAS.357.1197W,2008Natur.455..183R}.

Moreover, one needs to clarify that the
bimodal distribution of peak times for our sample can be affect by the selection effect. It is possible that the number of what we selected sample is a sun-class of total GRBs observed by Swift/XRT. On the other hand, the bimodal distribution of peak times is also possible affect by the sheltering from earth, namely, the dip of distribution is possible caused by the selection effect due to sheltered time from earth. Also, the derived parameters of early bumps (or late bumps) with magnetar (or black hole) fall-back model are dependent on the equation of state of neutron star (or initial mass of black hole) We hope that more observational data in X-ray band (i.e., Einstein probe) can be obtained in the future to identify the signature of central engine of GRBs.

\section*{Acknowledgements}
We acknowledge the use of the public data from the UK Swift Science Data Center. This work is supported by the Guangxi Science Foundation the National (grant No. 2023GXNSFDA026007), the Natural Science Foundation of China (grant Nos. 11922301 and 12133003), the Program of Bagui Scholars Program (LHJ), and the Guangxi Talent Program (“Highland of Innovation Talents”).

\section*{Data Availability}
This is the theoretical work, and there are no new data associated with this article. If one needs to adopt the data in this article, it should be cited this reference paper.

\bibliographystyle{mnras}
\bibliography{ms}
\clearpage
\begin{table*}
    
    \caption{The fitting results of our sample with smooth broken power-law model.}\label{tb1}
    \centering
    \setlength{\tabcolsep}{0.4cm}
    \begin{threeparttable}          
    \begin{tabular}{lcccccccc}\toprule
        GRB       &   T$_{90}$  &T$_{\rm start}$  &T$_{\rm end}$ &    $t_p$    &     $\alpha_1$       &     $\alpha_2 $    &      $z$  &      References\tnote{a} \\
        Name     &   (s)     &    (s)   &    (s) & (s)    &         &        &        &     \\ \hline
        Early bump\\
        \hline
        
          051016B   &   $4.0$   &   320 &   13500   &   $776_{-263}^{+703} $    &   $0.80 _{-0.05}^{+0.07} $    &   $ -1.86_{-0.31}^{+0.25}$    &   0.9364  & (1) \\
          060206    &   $7.0$   &   1200    &  12900    &   $3981_{-818}^{+1030}$    &    $1.26_{-0.04}^{+0.04}$    &    $-2.15_{0.82}^{-1.15}$    &    4.048  & (2)       \\
          070208 &   $48.0$ &   180 &   14000    & $954_{-63}^{+45} $     &    $1.52 _{-0.30}^{+0.21} $     &  $ -1.85_{-0.05}^{+0.01}$     &1.165 & (3)\\
          090429B & $5.5$   &   120 &   16800   &   $588_{-121}^{+262}$  &     $ 1.20 _{-0.12}^{+0.15}  $  &    $-1.36_{-1.21}^{+0.28}$     &9.4 & (4)\\
          091029 &   $39.2$ &   525 &   18900   & $1548_{-1202}^{+2524} $     &    $0.49 _{-0.06}^{+0.09} $     &  $ -1.71_{-0.15}^{+0.20}$     &2.752  &(5)\\
          120118B  & $23.3$   & 320 &   11000   & $1659_{-247}^{+290}$   &   $  1.01_{0.10}^{0.12}$       &    $ -1.21_{-0.39}^{+0.16}  $  &2.943 & (6)\\
          120213A &   $48.9$    &   620  &  12000   &  $3096_{-336} ^{+298} $  &     $2.41_{-0.32}^{+0.36} $     &    $-1.19_{-0.34}^{+0.14}$     &-&-\\
          120224A & $8.1$   &   220 &   19000   & $1071_{-220}^{+227}$     &     $0.78_{-0.06}^{+0.08}$      &    $-1.16_{-0.34}^{+0.12}$     &-&-\\
          121209A & $42.7$  &   100 &   7000    & $  707_{-76}^{+104} $  &      $1.23_{-0.06}^{+0.06}$    &   $-1.17_{-0.33}^{+0.13}$  &-&-\\
          140515A   & $23.4$   &    580 &   14800     &    $2691 _{-179}^{+470}$   &     $1.03_{-0.06}^{+0.08}$     &    $-2.90_{-0.61}^{+0.79} $&     6.32  &(7)         \\
          150911A & $7.2$   &   280 &   24800   &  $ 1479_{-330}^{+811}$   &     $ 1.29 _{-0.27}^{+0.64}$    &   $-1.49_{-1.05}^{+1.38}$     &-& -\\
          161129A & $35.53$    &    100 &   25000   &   $1412 _{-153}^{+172} $&     $ 2.02  _{-0.22}^{+0.22}$   &    $-1.07_{-0.18}^{+0.05} $    &0.645 &(8)\\
          170202A  &    $46.2$   &  300 &   59300   & $933_{-138}^{+163} $     &    $0.91 _{-0.04}^{+0.03} $     &  $ -1.31_{-0.35}^{+0.21}$     &3.645 & (9)\\
          170822A &$64.0$   &   500    &   19300    &   $2344_{-156}^{+167} $ &    $  1.72 _{-0.11}^{+0.10} $   &$-1.14_{-0.28}^{+0.10}$    &-&-\\
          181110A & $138.4$   &   400  &    36300   & $1659_{-247}^{+290} $   &     $1.72_{-0.10}^{+0.13}$      &    $-1.52_{-0.89}^{+0.39}  $   &1.505 & (10)\\
          190829A & $58.2$   &   500   &    56300   & $1380 _{-31}^{+65}$    &     $1.14 ^{-0.02}_{+0.02}$     &  $ -2.93_{-0.32}^{+0.31}$     &0.078 &(11)\\
          200917A & $19.4$    &     1000    &   36300   &  $2754_{-355}^{+793}$   &     $1.17_{-0.17}^{+0.20}$      &    $-2.55_{-0.97}^{+1.07} $     &-&-\\
          201128A & $5.41$    &  150    &   7300    & $457_{-131}^{+58}$   &     $0.80_{-0.07}^{+0.15}$      &    $-1.80_{-0.27}^{+1.02} $     &-&-\\
          220117A & $49.8$   &  350 &   13200   & $1380_{-121}^{+133} $     &    $1.55 _{-0.08}^{+0.09} $     &  $ -1.31_{-0.35}^{+0.21}$     &4.961 & (12)\\
\hline
Late bump\\
\hline
          071010A   & $6.0$      &  35000   &   533200       &   $ 52480^{+7775}_{-51433} $   &    $1.54_{-0.41}^{+0.32} $     &   $-2.16_{-1.11}^{+2.16} $   &    0.98   &(13)          \\
          081028      & $260.0$    &    9800    &   233200    &     $22387 _{-1007}^{+1055} $  &    $ 1.83_{-0.08}^{+0.08}  $  &     $-1.73_{-0.29}^{+0.25} $  &    3.038   &(14)           \\
          100901A   & $439.0$    &  9680    &   33320     &     $23442_{-1564}^{+1104}$   &     $1.35_{-0.07}^{+0.05}$     &   $ -1.12_{-0.23}^{+0.09} $  &   1.408   &(15)     \\
          120215A  & $26.5$     &   5160    &   53200   &   $ 10964_{-1632}^{+2217}$    &     $1.07_{-0.17}^{+0.24}$     & $-2.11_{-1.07}^{+0.71}$   &    -&    -\\
          120326A   & $69.6$    &   9560    &   330200   &     $35481_{-12039}^{+1672} $   &    $1.53_{-0.19}^{+0.11}$      &    $-1.11_{-1.00}^{+0.08}$    &    1.798     &(16)    \\
          130807A  &    37.7    &   3260    &     100200     &    $10471_{-2153}^{+2117} $   &     $1.21_{-0.22}^{+0.27} $    &     $-1.64_{-1.06}^{+0.44} $   &    -&    -\\
          131018A  &$73.2$&        4350    &   100000  & $7413_{-1387}^{+1707} $  &     $0.65_{-0.11}^{+0.12} $     &    $-1.88_{-1.43}^{+0.71}$     &    -&    -\\         
          150626B & $48.0$   &  5090    &   60000   &   $ 20892_{-1394}^{+1494}$     &   $  2.18 _{-0.17}^{+0.17}  $   &    $-1.32_{-0.43}^{+0.22}$    &   -&    -\\
          230414B & $25.9$   &  6490    &   69500   &    $21379_{-2758}^{+3167} $     &    $1.73 _{-0.13}^{+0.15} $     &  $ -1.27_{-0.03}^{+0.34}$     &3.568 & (17)\\\hline
          
      \end{tabular}
         \begin{tablenotes}    
        \footnotesize                      
        \item[a] The references of redshift for our sample. (1) \cite{2005GCN..4186....1S}; (2) \cite{2006A&A...451L..47F}; (3) \cite{2007GCN..6083....1C}; (4) \cite{2011ApJ...736....7C}; (5) \cite{2009GCN.10100....1C}; (6) \cite{2013GCN.14225....1M}; (7) \cite{2014GCN.16269....1C}; (8) \cite{2016GCN.20245....1C}; (9) \cite{2017GCN.20589....1P}; (10) \cite{2018GCN.23421....1P}; (11) \cite{2019GCN.25573....1L}; (12) \cite{2022GCN.31480....1P}; (13) \cite{2007GCN..6864....1P}; (14) \cite{2008GCN..8434....1B}; (15) \cite{2010GCN.11164....1C}; (16) \cite{2012GCN.13118....1T};  (17) \cite{2023GCN.33629....1A}.        
      \end{tablenotes}           
    \end{threeparttable}  
\end{table*}


 \begin{table}
    \caption{The MCMC fitting results of early bump sub-sample with fall-back accretion of magnetar}\label{tb2}
    \centering
    \setlength{\tabcolsep}{0.55cm}
    \begin{threeparttable}          
    \begin{tabular}{lcccc}\toprule
        GRB\tnote{a}& ${\rm log~t_1}$ & ${\rm log~\eta_{\rm mag}}$ & $\rm log~(B_0)$ &  $P_0$ \\ 
        Name& (s) && (G) &  (ms) \\ 
        \hline

          051016B   &  $2.60_{-0.11}^{+0.11}$ & $-2.06_{-0.22}^{+0.18}$ &   $13.69_{-0.04}^{+0.04}$ &   $1.11_{-0.12}^{+0.13}$  \\
          060206    &$3.49_{-0.02}^{+0.02}$ &   $-1.88_{-1.88}^{+0.02}$ &   $13.72_{-0.02}^{+0.03}$ &   $1.48_{-0.35}^{+0.34}$  \\
          070208    &   $2.13_{-0.09}^{+0.12}$   &   $-1.12_{-0.23}^{+0.16}$  &   $13.74_{-0.02}^{+0.02}$  &   $1.34_{-0.20}^{+0.30}$    \\
			090429B    &   $2.67_{-0.07}^{+0.08}$    & $ -0.77_{-0.07}^{+0.07}$    &   $14.00_{-0.05}^{+0.04}$  &    $0.57_{-0.05}^{+0.10}$    \\
          091029    &   $3.26_{-0.04}^{+0.06}$  &   $-1.88_{-0.05}^{+0.04}$ &   $13.73_{-0.03}^{+0.03}$ &   $15.42_{-6.71}^{+10.33}$    \\
          120118B    &   $2.58_{-0.13}^{+0.21}$    &$ -1.83_{-0.43}^{+0.24}$   &   $13.60_{-0.03}^{+0.03}$  &    $1.19_{-0.18}^{+0.25}$    \\
          120213A$^{\ast}$   &    $3.36_{-0.03}^{+0.03}$   &   $-2.16_{-0.04}^{+0.04}$   &   $13.96_{-0.06}^{+0.03}$  &    $1.33_{-0.32}^{+0.45}$    \\
          120224A$^{\ast}$   &   $2.76_{-0.20}^{+0.24}$    &   $-2.03_{-0.31}^{+0.31}$   &   $13.58_{-0.03}^{+0.03}$   &   $0.82_{-0.10}^{+0.09}$    \\
          121209A$^{\ast}$   &   $2.15_{-0.12}^{+0.28}$    &   $-1.05_{-0.51}^{+0.34} $  &   $13.78_{-0.05}^{+0.03}$   &   $0.80_{-0.13}^{+0.42}$    \\
          140515A   &   $3.50_{-0.04}^{+0.04}$  &   $-1.81_{-0.04}^{+0.03}$ &   $13.82_{-0.05}^{+0.04}$ &   $1.85_{-0.75}^{+0.77}$  \\
          150911A$^{\ast}$   &   $2.69_{-0.05}^{+0.06}$    &   $-1.51_{-0.07}^{+0.06} $  &   $13.80_{-0.01}^{+0.01}$   &  $ 8.49_{-3.91}^{+4.36} $   \\
          161129A   &   $3.04_{-0.04}^{+0.03}$    &   $-1.76_{-0.03}^{+0.03}  $&    $13.96_{-0.05}^{+0.03}$   & $  0.90_{-0.05}^{+0.06} $   \\
          170202A   &   $2.74_{-0.04}^{+0.05}$    &   $-1.20_{-0.06}^{+0.06}$   &   $13.67_{-0.03}^{+0.044}$    &   $0.47_{-0.06}^{+0.10}$  \\
          170822A$^{\ast}$   &   $3.13 _{-0.05}^{+0.07} $  &   $-1.87_{-0.02}^{+0.02} $  &   $13.78_{-0.04}^{+0.07}$   &  $ 0.98_{-0.14}^{+0.27} $   \\
          181110A   &   $3.08_{-0.06}^{+0.11}$    &   $-1.55_{-0.04}^{+0.03}$   &   $13.92_{-0.04}^{+0.17} $  &  $ 1.62_{-0.40}^{+0.28}  $  \\
	190829A   &   $2.92_{-0.02}^{+0.02}$    &  $ -2.21_{-0.02}^{+0.01} $  &   $13.86_{-0.01}^{+0.02}$   &   $9.75_{-0.44}^{+0.32}$    \\
          200917A$^{\ast}$   &   $3.02_{-0.08}^{+0.09}$  &   $-2.09_{-0.09}^{+0.08}$ &   $13.61_{-0.04}^{+0.04}$ &   $15.13_{-8.69}^{+9.88}$ \\
          201128A$^{\ast}$   &   $2.32_{-0.11}^{+0.12}$  &   $-1.43_{-0.19}^{+0.15}$ &   $13.82_{-0.04}^{+0.02}$ &   $1.20_{-0.15}^{+0.15}$  \\
          220117A   &   $2.74_{-0.03}^{+0.07} $   &  $ -0.92_{-0.07}^{+0.05}$   &   $13.83_{-0.03}^{+0.04}$   &   $0.58_{-0.06}^{+0.53}$    \\\hline
      \end{tabular}
         \begin{tablenotes}    
        \footnotesize               
        \item[a] No redshift measured of our sub-sample are mark with $[\ast]$, and we adopt $z=1$ to do the MCMC fits.
      \end{tablenotes}            
    \end{threeparttable}       
  \end{table}
 \begin{table}
    \caption{The MCMC fitting and calculated results of late bump sub-sample with fall-back accretion of black hole}\label{tb3}
    \centering
    \setlength{\tabcolsep}{0.4cm}
    \begin{threeparttable}          
      \begin{tabular}{lccccc}\toprule
        GRB& $\rm log(T_{p})$ &  s & log$(\dot{m}_{p})$& $\rm  log(M_{acc})$& $R_{\rm fb}$ \\ 
        Name& (s) &  &($\rm M_{\odot}~s^{-1}$)&$\rm (M_\odot)$& ($10^{11}$ cm) \\
        \hline
			
          071010A&$4.88_{-0.02}^{+0.03}$  & $1.48_{-0.57}^{+0.65}$  &   $-7.77_{-0.08}^{+0.06}$ &   $-2.99$&4.23 \\
          081028  &$ 4.44_{-0.01}^{+0.01}$ & $1.22_{-0.19}^{+0.25}$ &$-5.79_{-0.03}^{+0.03}$& $-1.75$&1.21 \\
          100901A&$4.52_{-0.01}^{+0.01}$&$0.77_{-0.10}^{+0.12}$&$-6.51_{-0.03}^{+0.03}$&$-2.05$&1.74\\
          120215A$^{\ast}$&$4.30_{-0.04}^{+0.04}$& $0.87_{-0.23}^{+0.36}$ & $-7.69_{-0.06}^{+0.05} $ & $-3.38$&1.26\\
          120326A&$4.67_{-0.02}^{+0.03}$&$0.84_{-0.16}^{+0.21}$&$-6.24_{-0.03}^{+0.03}$&$-1.85$&1.93\\
          130807A$^{\ast}$   &   $4.26_{-0.05}^{+0.06}$  &   $1.12_{-0.38}^{+0.67}$ &   $-7.42_{-0.08}^{+0.08}$ &   $-3.14$&1.04\\
          131018A$^{\ast}$&$4.47_{-0.07}^{+0.08}$&$0.33_{-0.07}^{+0.10}$&$-7.48_{-0.06}^{+0.05}$&$-2.86$& 1.19\\
          150626B$^{\ast}$&$4.35_{-0.02}^{+0.02}$&$1.53_{-0.36}^{+0.52}$&$-6.74_{-0.03}^{+0.03}$&$-2.62$&2.02\\
          230414B&$4.49_{-0.04}^{+0.06}$&$0.64_{-0.22}^{+0.30}$&$-7.38_{-0.11}^{+0.09}$&$-3.38$& 1.09\\
          \hline
      \end{tabular}
         \begin{tablenotes}    
        \footnotesize               
        \item[a] No redshift measured of our sub-sample are mark with $[\ast]$, and we adopt $z=1$ to do the MCMC fits and calculations.
      \end{tablenotes}            
    \end{threeparttable}       
  \end{table}
\clearpage
 \begin{figure*}
    \includegraphics[width=3.5in]{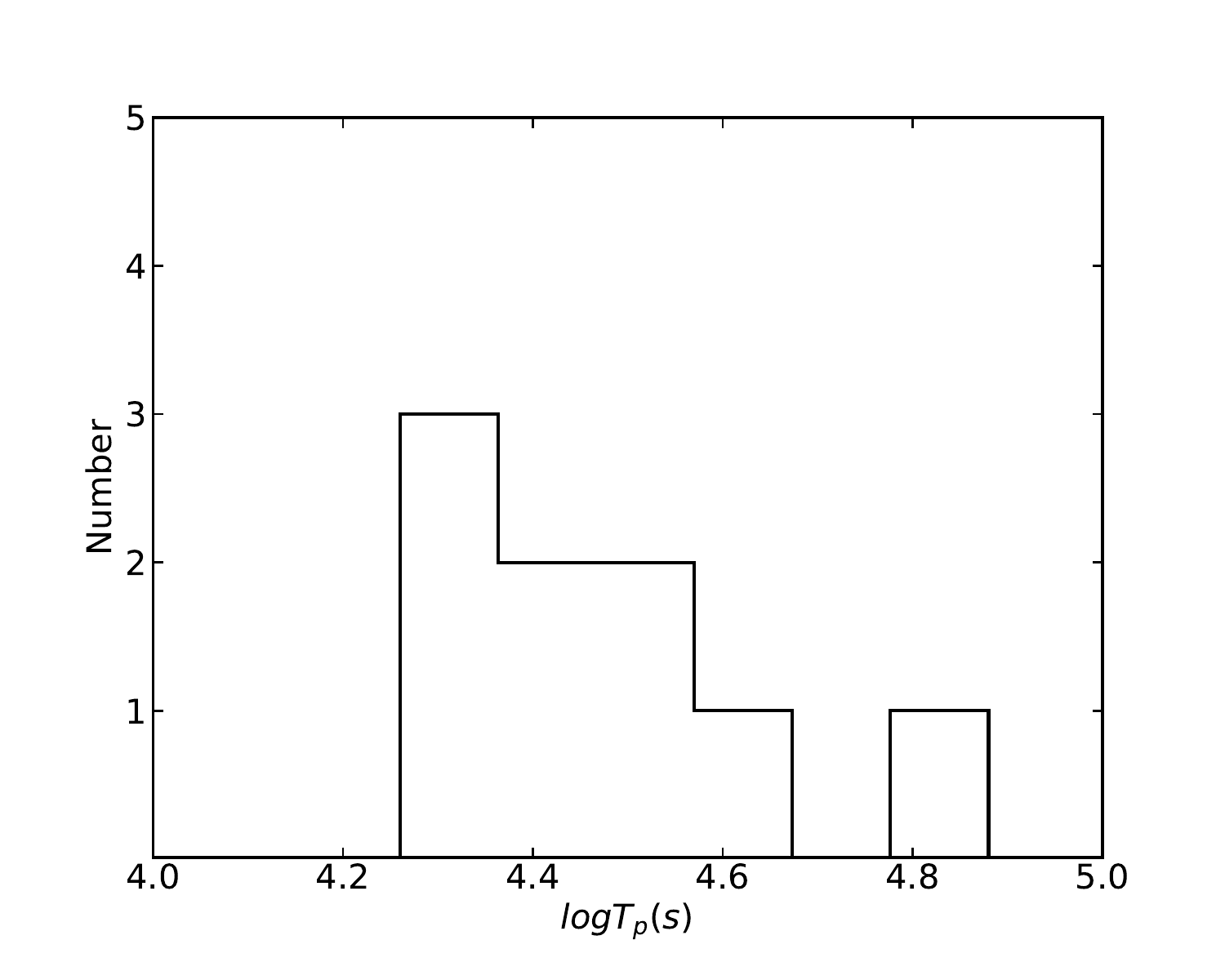}
    \includegraphics[width=3.5in]{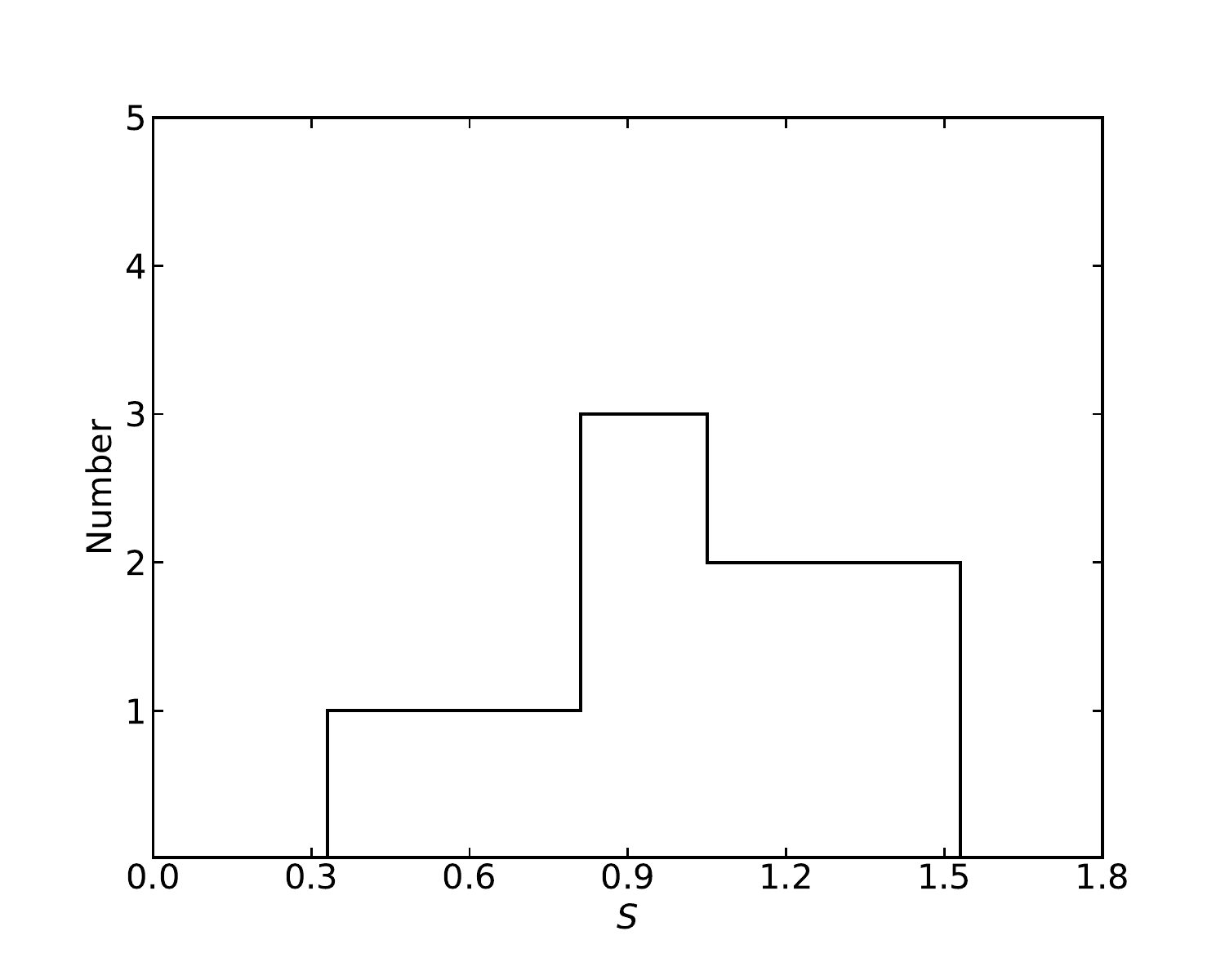}
    \includegraphics[width=3.5in]{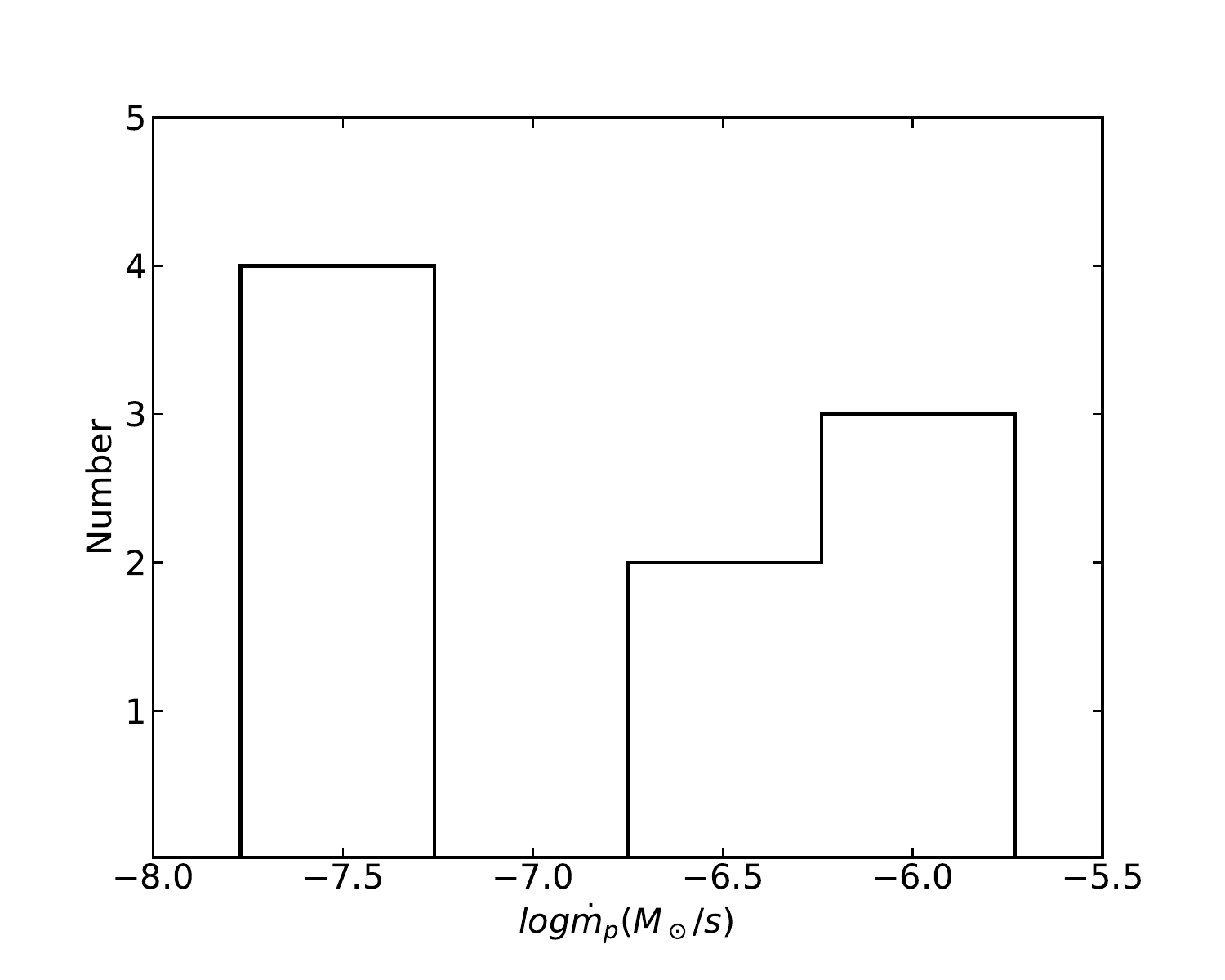}
    \includegraphics[width=3.5in]{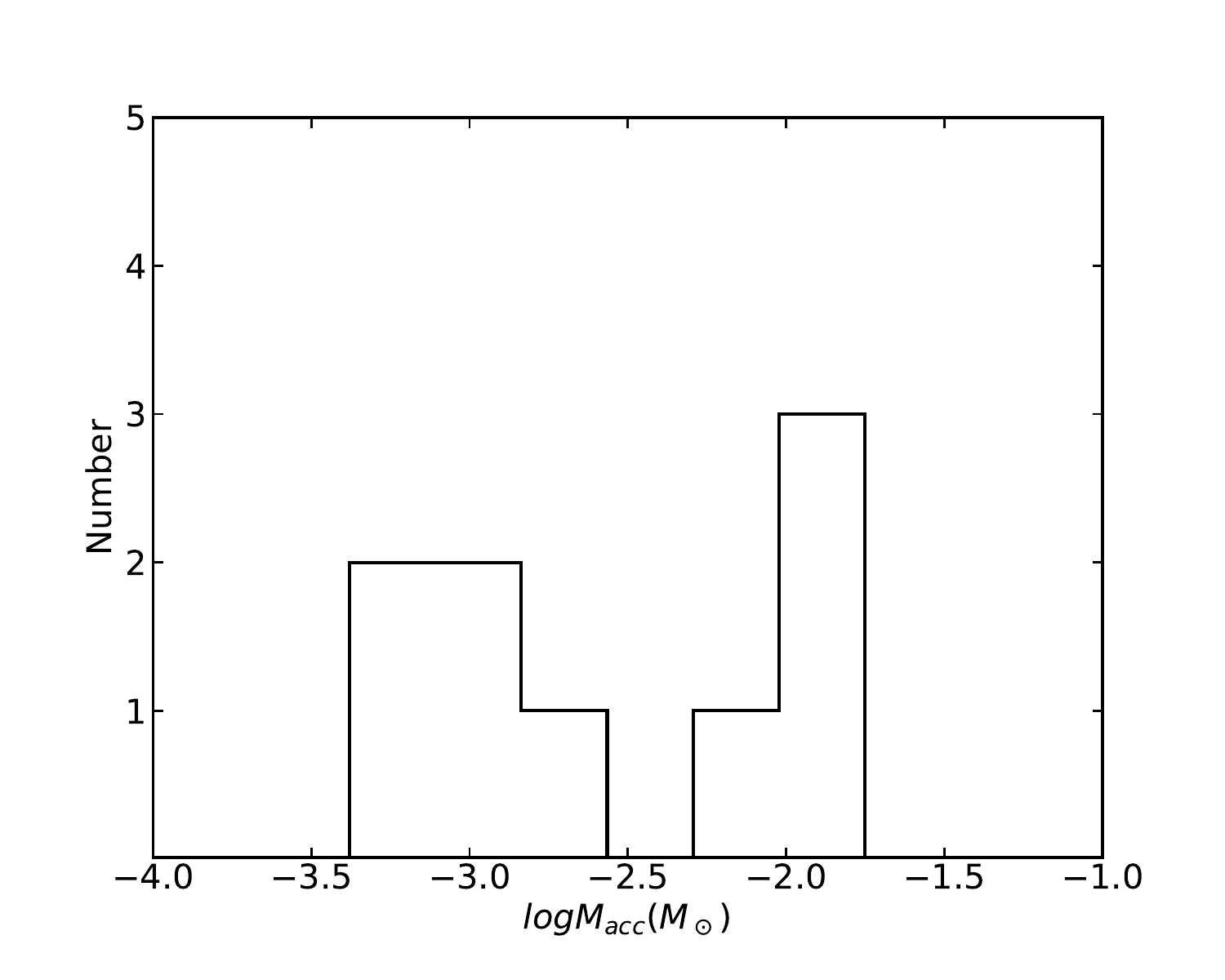}
    \includegraphics[width=3.5in]{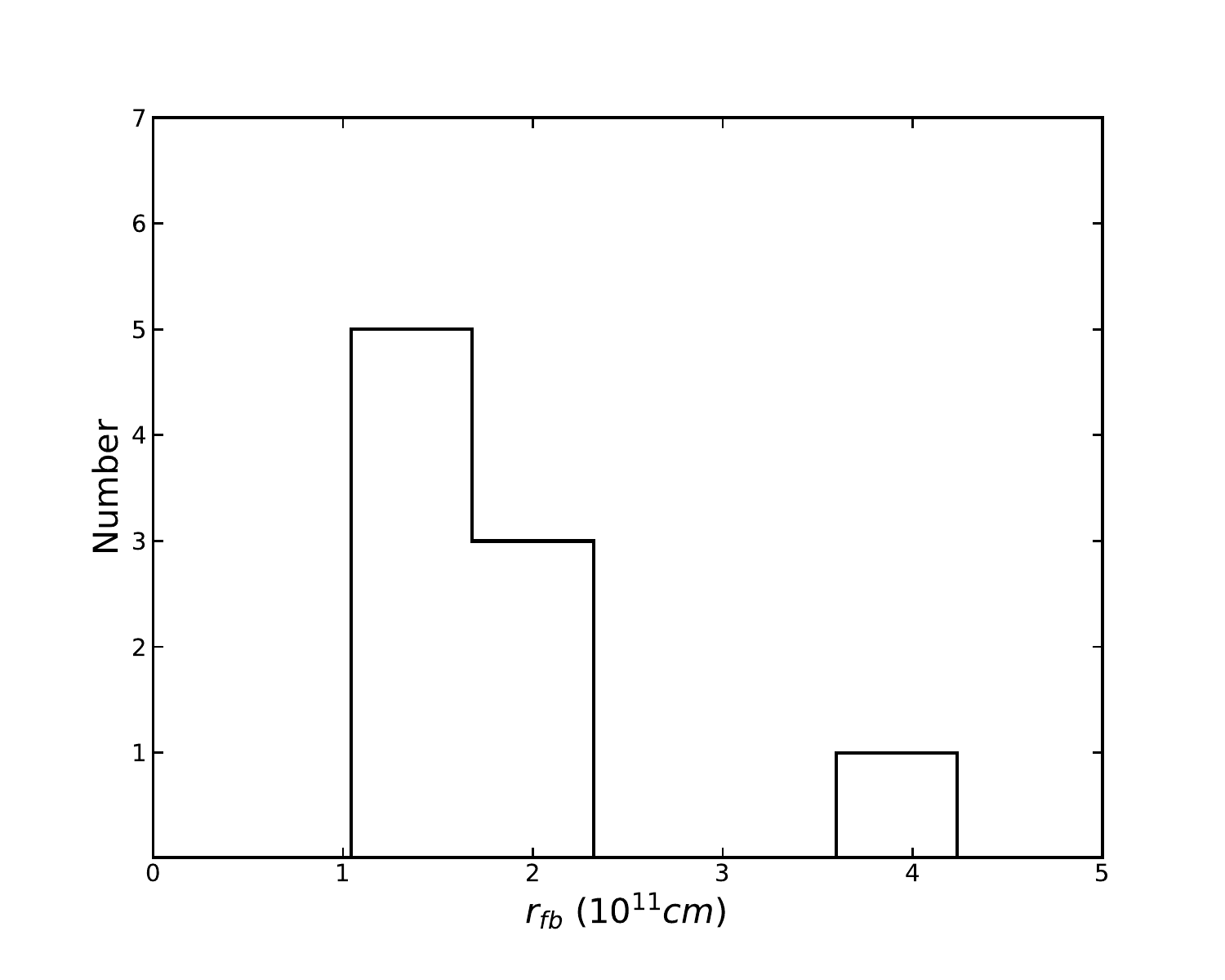}
    \caption{Distributions of free parameters for late bumps with fall-back accretion of black hole model.}
        \label{Fig5}
\end{figure*}


\end{document}